\documentclass[sigconf]{aamas}
\usepackage[ruled,vlined, noend]{algorithm2e}
\SetKwComment{Comment}{$\backslash\backslash$\ }{}

\usepackage{tikz}
\usetikzlibrary{automata, positioning, arrows}
\tikzset{
->, % makes the edges directed
>=stealth, % makes the arrow heads bold
node distance=2.2cm, % specifies the minimum distance between two nodes. Change if necessary.
initial text=$ $, % sets the text that appears on the start arrow
}
\usepackage{balance} % for balancing columns on the final page
\usepackage{soul}
\usepackage[normalem]{ulem}
\usepackage[shortlabels]{enumitem}
\usepackage{bm}

\acmSubmissionID{44}

\title{An Abstraction-based Method to Check\\ Multi-Agent Deep Reinforcement-Learning Behaviors\\ (Extended Version)}

\author{Pierre El Mqirmi}
\affiliation{
  \institution{Imperial College London, UK}}
\email{pierre.el-mqirmi19@imperial.ac.uk}

\author{Francesco Belardinelli}
\affiliation{
  \institution{Imperial College London, UK\\
  Universit\'e d'Evry, France}}
\email{francesco.belardinelli@imperial.ac.uk}

\author{Borja G. León} %Please when we publish by names, use this
\affiliation{
  \institution{Imperial College London, UK}}
\email{b.gonzalez-leon19@imperial.ac.uk}

\begin{abstract}
Multi-agent reinforcement learning (RL) often struggles to ensure the safe behaviours of the learning agents, and therefore it is generally not adapted to safety-critical applications. To address this issue, we present a methodology that combines formal verification with (deep) RL algorithms to guarantee the satisfaction of formally-specified safety constraints both in training and testing. The approach we propose expresses the constraints to verify in \textit{Probabilistic Computation Tree Logic} (PCTL) and builds an abstract representation of the system to reduce the complexity of the verification step. This abstract model allows for model checking techniques to identify a set of abstract policies that meet the safety constraints expressed in PCTL. Then, the agents' behaviours are restricted according to these safe abstract policies.
We provide formal guarantees that by using this method, the actions of the agents always meet the safety constraints, and provide a procedure to generate an abstract model automatically. We empirically evaluate and show the effectiveness of our method in a multi-agent environment.
\end{abstract}


\newcommand{\BibTeX}{\rm B\kern-.05em{\sc i\kern-.025em b}\kern-.08em\TeX}

\begin{document}

%%% The following commands remove the headers in your paper. For final 
%%% papers, these will be inserted during the pagination process.

\pagestyle{fancy}
\fancyhead{}

%%% The next command prints the information defined in the preamble.

\maketitle

%%%%%%%%%%%%%%%%%%%%%%%%%%%%%%%%%%%%%%%%%%%%%%%%%%%%%%%%%%%%%%%%%%%%%%%%

\section{Introduction}

Autonomous agents acting in unknown environments are attracting research interests due to their potential applications in multiple domains including robotics, network optimisation, and distributed resource allocation \cite{yang2004multiagent, pandey2016adaptive, zhang2009multi, wiering2000multi, riedmiller2000reinforcement}. Currently, one of the most popular techniques to tackle these domains is reinforcement learning (RL) \cite{RLIntro}. However, in order to learn how to act, RL requires to explore the environment, which in safety-critical scenarios means that the agents will commonly take dangerous actions, possibly damaging themselves or even putting humans at risk. Consequently, RL and its extension deep RL (DRL) \cite{IntroDRL} are rarely used in real-world applications where multiple safety-critical constraints need to be satisfied simultaneously. To alleviate this problem, (D)RL algorithms are being combined with formal verification techniques to ensure safety in learning.
%and policies that satisfy these constraints. 
Even though significant progress has been achieved in this direction \cite{AssuranceInReinforcement,safeRLviaShielding,shieldedDecisionMakinginMDP,VerificationRepairControlPolicies,safeRL,cheng2019end}, settings with multiple learning agents are comparatively less explored and understood.

\paragraph{Our Contribution}
In this paper we introduce {\em assured multi-agent reinforcement learning} (AMARL), a method to formally guarantee the safe behaviour of agents acting in an unknown environment through the satisfaction of safety constraints
by the solution learned using a DRL algorithm, both at training and test time. Building upon
%Taking as a baseline 
the \textit{assured reinforcement learning} (ARL) technique 
%proposed by Mason et Al. 
in \cite{AssuranceInReinforcement}, we combine reinforcement learning and formal verification \cite{QV} 
%to address the limitations of traditional (D)RL algorithms and 
to ensure the satisfaction of constraints expressed in \textit{Probabilistic Computation Tree Logic} (PCTL) \cite{PCTL}. Differently from ARL, we support a multi-agent setting and 
%the use of 
DRL algorithms.
%\fb{we might say which ones} \pe{Shouldn't we stay more general here?} \bg{I would just say DRL here}. 
Specifically, we introduce the notion of \textit{abstract Markov game} (AMG) 
%as a multi-agent extension of Markov decision processes (MDP) \pe{Do you mean Abstract MDP here? (AMDP)}, 
and present a procedure to generate AMGs automatically, unlike ARL where abstract models are handcrafted. Moreover, 
%In contrast to \textit{ARL}, 
we provide formal proofs of the preservation of properties expressed in (fragments of) PCTL between the abstract and concrete model. %\sout{and ensure the satisfaction of the safety constraints by the mean of a \textit{shield} \pe{, i.e. an entity that monitors the agents' actions.} \cite{safeRLviaShielding, shieldedDecisionMakinginMDP}. 
%\fb{please add relevant ref for shield}.}

Multiple challenges arise from multi-agent settings, such as the {\em curse of dimensionality} \cite{MARL}. It is therefore crucial to build a small enough abstract Markov game while preserving all the required properties. Moreover, many RL algorithms cannot guarantee convergence in multi-agents scenarios and are therefore harder to train. Finally, the use of \textit{options} (i.e., temporally extended actions) in the abstract MG makes the definition of the reward and transition functions complex.
%given that the options of the agents may
Nonetheless, our experimental results demonstrate the effectiveness of the AMARL method to ensure safety constraints. 
%\sout{and to learn to complete the selected abstract joint policy in the concrete model.}
%show that multi-agent solutions can sometimes surpass single-agent solutions 
%\bg{I don't remember this when reading Section 4, where do we prove this?} \pe{I think we removed this part but the point was that the solution learned with multiple agents out perform the solution of a single agent found in the reference paper. However since we don't have the space to show that, I think we should remove this sentence} \bg{I agree to remove it then}. 
Moreover, we demonstrate its compatibility with DRL algorithms and its ability to ensure the safe behaviours of agents even during the learning stage.
%%%%%%%%%%%%%%%%%%%%%%%%%%%%%%%%%%%%%%%%%%%%%%%%%%%%%%%%%%%%%%%%%%%%%%%%

\paragraph{Related Work} \label{section:relatedWork}
This paper builds upon the ARL method introduced in \cite{AssuranceInReinforcement}; consequently, both our method and ARL are closely related and belong to the same class of safe RL techniques based on restricting 
%the adjustment of the
exploration \cite{safeRL}. Further, they both 
%\textit{ARL} and our method 
support constraints expressed in the probabilistic temporal logic PCTL. Nevertheless,
our AMARL method is more general than ARL,
as 
%} \cite{AssuranceInReinforcement} in the sense that 
we support both
%we extend \textit{ARL} to 
multi-agent settings and 
%we support 
the use of both tabular RL and DRL. Moreover, in contrast with ARL, the abstract representation of the Markov game is built in an automated manner and proofs of constraint preservation between abstracts and concrete models are provided. Hence, ARL can be seen as a special case of our method, where only single-agent tabular RL is supported and no proof of constraint preservation 
%between the manually-built abstract model and the concrete model 
is provided. 
%\sout{Finally, the last difference w.r.t.~\textit{ARL} is the use of a {\em shield} to prevent unsafe actions \pe{, i.e. actions that are not part of the selected abstract joint-policy,} instead of removing actions from the action space of the agents.}

Our approach differs from other safe RL methods based on formal verification, as it relies on the construction of an abstract model to tackle the high dimensional spaces found in typical {\em shield} methods \cite{safeRLviaShielding, shieldedDecisionMakinginMDP} and allows to find high-level solutions directly from the abstract model.
The approach proposed 
%by Alshiekh et al. 
in \cite{safeRLviaShielding} introduced the notion of shield, i.e., an entity that monitors the agents' actions, and also expresses constraints in temporal logic. A major difference w.r.t.~AMARL is that their shield instead of penalizing unsafe actions and preventing the agent from interacting with the environment, replaces the unsafe action with another safe action and therefore requires the shield to be active both at training and test time. Moreover, their method is designed for single-agent RL and constraints are expressed in linear temporal logic \cite{pnueli1977temporal}. 
%Thus, unlike our technique, their method does not support probabilistic and multi-agent environments.
%
Similarly to \cite{safeRLviaShielding} and AMARL, 
%Jansen et al. 
\cite{shieldedDecisionMakinginMDP} proposes a shield-based technique that expresses constraints in PCTL. A key difference here is that they consider multi-agent settings where only a single agent is controllable. Their method also relies on the construction of an abstraction. However, differently from AMARL, their abstract model does not include the reward function, thus preventing from solving the problem at the abstract level.
The approach presented in \cite{VerificationRepairControlPolicies} also expresses safety constraints in PCTL, but instead of ensuring safety by mean of a shield, this method is based on verification and repair of the learned policy. Hence, it does not provide any safety guarantee at learning time.
%as 
%they repair 
%the learnt policy is repaired 
%as it is learned accordingly to some 
%based on the counter-examples returned by the model checker. 
Moreover, a significant limitation of \cite{VerificationRepairControlPolicies} is its difficulty to tackle high dimensional state spaces. Thus making this method prone
to the curse of dimensionality and not adapted for multi-agent settings that are known to suffer from this problem.

\section{Background}

In this section, we provide the necessary background regarding both reinforcement learning and formal verification. 
%\sout{Moreover, we present the \textit{assured reinforcement learning} (ARL) method introduced in \cite{AssuranceInReinforcement}, whereupon we build in in this work.}
%This chapter seeks to offer all the knowledge required to understand our approach and the contributions we are making.

\subsection{Multi-agent Reinforcement Learning}

Multi-agent reinforcement learning (MARL) 
%\bg{use acronyms (MARL)} \pe{Don't we need to write it completely at least one time?} 
is a machine learning \cite{bishop2006pattern} technique where agents situated in an environment aim to maximise an expected reward signal provided by their interactions with such environment \cite{RLIntro}. This problem is typically modelled as a Markov game (MG). Initially, this model is unknown to the agents and they explore it to collect rewards and to learn the optimal behaviours (i.e., policies). %\sout{We therefore introduce Markov games to model multi-agent reinforcement learning (MARL) problems.}
\begin{definition}[Markov Game] \cite{markovGame}
    A {\em Markov game} with $n$-agents is a tuple $M = \langle S, A_1, \ldots , A_n,$ $P, R_1, \ldots , R_n, \gamma \rangle$ where:
    \begin{itemize}
        \item $S$ is the {\em state space}.
        \item For every $i \leq n$, $A_i$ is the {\em action space} of agent $i$.
        \item $P$ is the {\em transition function} where $P^{\vec{a}}_{ss'}$ denotes the probability of going from state $s \in S$ to state $s' \in S$ by taking the joint action $\vec{a}= \langle a_1, \ldots, a_n \rangle$, where 
        %In the joint action $\vec{a} = \langle a_1, \ldots , a_n \rangle$, 
        all $a_i \in A_i$ are the actions taken by the agents simultaneously.
        \item $R_1, \ldots , R_n$ is a set of \textit{reward functions}, where $R^{\vec{\rm a}}_{i, ss'}$ denotes the reward received by agent $i$ when the joint action $\vec{a}$ from state $s \in S$ to state $s' \in S$ is performed. We denote $r_{i,t}$ the reward received by agent $i$ at time step $t$.
        \item $\gamma \in [0, 1]$ is the discount factor.
%        \item $n$ is the number of agents.
    \end{itemize}
\end{definition}
In MARL each agent's behaviour is determined by her policy. In this work, we use deterministic policies and define the policy of agent $i$ as $\pi_i: S \rightarrow A_i$. Each agent $i$ tries to find an optimal policy $\pi^*_i$ that maximises the sum $R_i = \sum_{k=0}^\infty \gamma^k r_{i, t+k}$ of her expected rewards. Finally, a {\em joint policy} is a tuple $\vec{\pi} = \langle \pi_1, \ldots , \pi_n \rangle$.
%The tuple of policies $\langle \pi_1, \ldots , \pi_n \rangle$ form collectively the {\em joint policy} $\vec{\pi}$. 
% : $\sum_{k=0}^\infty \gamma^k r_{i, t+k}$ \cite{MARL}. A (stochastic) {\em policy} $\pi_i : S \times A_i \rightarrow [0, 1]$ \bg{Note that you used DQN and this algorithm uses a deterministic policy at test time} is a distribution over state-action pairs in  $S \times A_i$.
% Finally, a {\em joint policy} is a tuple $\vec{\pi} = \langle \pi_1, \ldots , \pi_n \rangle$.

%The autonomous agents interacting with the environment can have cooperative, competitive, or independent behaviours \cite{MARL}. However, 
Moreover, 
%at the abstract level, 
we focus on fully cooperative problems \cite{MARL}, where all the agents share the same reward function $R_1 = \ldots  = R_n$ and try to maximise their common sum of rewards. 
%\pe{I am unsure if it is clear that we focus on fully cooperative problems but only with a high-level point of view. I.e. we use multiple agents with a single objective but the agents can be trained with different reward functions. Also, maybe it is sufficient to just explain the three type of Markov game and not specifying now on which model we focus as we explain it later.}
% Therefore, according to \cite{MARL}, we can identify three different types of Markov game: \bg{I think this could be omitted and focus only on cooperative games, which are the ones we have worked with}
% \begin{itemize}
%     \item \textbf{Fully cooperative}: all agents share the same reward function $R_1 = \ldots  = R_n$ and try to maximise their common sum of reward.
%     %Thus, we have that .
%     \item \textbf{Fully competitive}: two agents (i.e., $n=2$) who have opposite goals. The reward function $R_1$ of agent $1$ is the negation of the reward function $R_2$ of agent $2$.
%     \item \textbf{Mixed}: neither fully cooperative nor fully competitive, each agent has its own reward function, and each agent tries to maximise their own goal.
% \end{itemize}
The method we propose makes use of the popular Independent Q-Learning (IQL)  algorithm \cite{IndependentQ} due to its simplicity and to the fact that it is the direct extension of the tabular Q-Learning typically used in previous single-agent shielded based approaches \cite{AssuranceInReinforcement, safeRLviaShielding, shieldedDecisionMakinginMDP}. 
%\bg{let's try not to sell ourselves as s simple extension of ARL, cite here "Safe Reinforcement Learning via Shielding" and "Shielded Decision-Making in MDPs" as well}. 
According to IQL, each agent $i$ has her own Q-table and update the values of the state-action pairs 
%of its Q-table 
as follows:
\begin{equation}
    Q_{i, t+1}(s, a_i) \leftarrow Q_{i, t}(s, a_i) + \alpha [r_{i, t} + \gamma \cdot \max_{a_i' \in A_i} Q_{i, t}(s', a_i') - Q_{i, t}(s, a_i)]
\end{equation}
where $0 < \alpha \leq 1$ is the learning rate. Once the learning phase is over, the optimal policy of an agent using IQL returns the action with the highest Q-value according to the state-action pairs of her Q-table. In this paper, we make use of deep RL, which combines RL and deep learning \cite{goodfellow2016deep} %\bg{cite the book of DL from Goodfellow} 
to increase the scalability of traditional tabular RL algorithms \cite{mnih2015human}. 
%\bg{you can cite the DQN paper "human-level performance..."}.
In particular, we use the direct extension of IQL called independent deep Q-Learning (IDQL) \cite{IDQN}, where the Q-tables of the agents are replaced by neural networks.

%\sout{Then, single-agent reinforcement learning can be seen as a special case of MARL with a single agent. The system is then represented as a Markov decision process (MDP), i.e., a Markov game with $n = 1$}. 
%\bg{We need to add one/two lines explaining what is DRL since currently we only say what is MARL and RL}

\paragraph{Shield in Safe RL}

Our approach makes use of a {\em shield} \cite{shieldedDecisionMakinginMDP, safeRLviaShielding, cheng2019end} to ensure the fulfillment of safety constraints both in training and testing. A shield is an entity that monitors the agents' actions and prevents them from performing any action that would lead to an {\em unsafe state}. Therefore, the shield is generally seen as an intermediate between the agents and the environment that ensure that only {\em safe actions} are performed on the environment.
%\bg{We should say here what is exactly an unsafe state since in literature they take different approaches, i.e., is it a state that directly violates a safety constrain? or any state acting as a non-return point from which the agent can't avoid the future violation of the specification?}. 
Note that it is desired to have a shield that intervenes as little as possible and that does not impact too much the agents' exploration freedom. Otherwise, the shield could prevent the agents from finding optimal policies. For this reason, it is common to have a shield that penalizes the agents with a negative reward when intervening.

%\pe{I am not sure what to say more about the shield... Do you have any suggestion Borja?}
%\fb{we could explain how the shield is implemented in our particular case, possibly providing references.}\bg{we should explain how does the shield interacts within the RL loop, e.g., what does the agent 'observe'?, what kind of reward is received when the shield intervenes? I know in some works they give a penalisation but I think it's not our case. Additionally, in our experimental section, it would be nice to indicate how many times the shield has to intervene at test time}

\subsection{Formal Verification by Model Checking}

The method we propose makes use of the \textit{probabilistic computation-tree temporal logic} (PCTL) \cite{PCTL} to express the constraints to satisfy, possibly
%Moreover, since we work with rewards, we use \textit{PCTL} 
extended with rewards \cite{sfm11}. 
%We first provide the syntax and semantics of PCTL and then present its extension that allows us to include the verification of rewards in the model.

%\subsubsection{Probabilistic Computation-tree Temporal Logic}

%\paragraph{Syntax.} 
%The syntax of PCTL is defined in terms of \textit{state} and \textit{path formulae}. 
We  consider a set $AP$ of \textit{atomic propositions} (or atoms) to label the states of an MG, in order to express that some facts hold at certain states \cite{principlesmodelchecking}.

\begin{definition}[PCTL] \label{def:pctl}
Formulae $\Phi$ over a set $AP$ of atoms are built according to the following grammar:
%\cite{principlesmodelchecking, sfm11}:
\begin{eqnarray*}
    \Phi & ::= & true \mid a \mid \Phi \land \Phi \mid \neg \Phi \mid P_{\sim p}(\varphi)\\
    \varphi & ::= & \bigcirc \Phi \mid \Phi U \Phi \mid \Phi U^{\leq n} \Phi
\end{eqnarray*}
where $a \in AP$, $\varphi$ is a path formula, $\sim$ $\in \{<, \leq, >, \geq\}$, $p \in [0, 1]$, 
%$\Phi$, $\Phi_1$ and $\Phi_2$ are state formulae, 
and $n \in \mathbb{N}$. 
Intuitively, {\em next } $\bigcirc$, {\em until} $U$, and {\em bounded until} $U^{\leq n}$ are the standard temporal operators; while 
%\begin{itemize}
%    \item $\Phi$, $\Phi_1$ and $\Phi_2$ are state formulae.
%    \item $n \in \mathbb{N}$.
%    \item $\bigcirc$ is the \textit{next operator} meaning that formula $\bigcirc \Phi$ holds iff $\Phi$ holds at the next state in the path.
%    \item $U$ is the \textit{until operator}: $\Phi_1 U \Phi_2$ means that eventually $\Phi_2$ will be true and all previous states in the path satisfy $\Phi_1$.
%    \item $U^{\leq n}$ is the bounded version of the \textit{until operator}. %$\Phi_1 U^{\leq n} \Phi_2$ means that before $\Phi_2$ becomes true, previous states $s_0,\ldots , s_k$, $k < n$ in the path satisfy $\Phi_1$. 
%    \item 
$P_{\sim p}(\varphi)$ expresses that the path formula $\varphi$ is true with  probability $\sim p$.
%on all paths starting from the current state.
%\end{itemize}
\end{definition}
PCTL formulae are interpreted over the states and paths of a Markov game and $s \models \Phi$ denotes that state formula $\Phi$ holds in state $s$.
%
%\fb{the semantics is not really provided, we might put it in the supplementary material, if allowed.}
%
%\paragraph{Semantics.} 
In particular, for an MG $M$, we introduce a labelling function $L: S \to 2^{AP}$.
%as the  that maps each state $s \in S$ of $M$ to a set of atomic propositions $AP$.
For reasons of space, we omit the details about the semantics of PCTL and refer to \cite{principlesmodelchecking, sfm11}.
%for details.

%\subsubsection{PCTL extended with rewards in PRISM}

%\begin{definition}
The extended version of PCTL with rewards supported by the PRISM language \cite{PRISM} extends the definition of state formulae with the following clauses \cite{sfm11}: 
\begin{equation*}
    \Phi ::= R_{\sim r}(I^{=k}) \mid R_{\sim r}(C^{\leq k}) \mid R_{\sim r}(\lozenge \Phi)
\end{equation*}
where $\sim$ $\in \{<, \leq, >, \geq\}$ and $r \in \mathbb{R}_{\geq 0}$.

Intuitively, $R_{\sim r}(I^{=k})$ expresses the reward at time step $k$, $R_{\sim r}(C^{\leq k})$ expresses the expected cumulative reward up to time $k$, and $R_{\sim r}(\lozenge \Phi)$ expresses the expected cumulative reward to reach a state that satisfies $\Phi$.
%\end{definition}

Finally, we introduce the {\em weak fragment of PCTL} (or wPCTL) that discards the next and bounded until operators. That is, 
%state formulae in wPCTL are defined as in PCTL, whereas 
path formulae in wPCTL are restricted as follows:
\begin{eqnarray*}
\varphi & ::= & \Phi U \Phi
\end{eqnarray*}
The weak fragment of PCTL features preeminently as the language to express constraint on learning agents, including those stated in Table \ref{safety-constraints}. The Storm model checker  \cite{Storm}  supports the same specification language as PRISM, and we use it to verify the satisfaction of such constraints
%hold in an MG. %\fb{which model checker do we use? -- We use Storm but with the PRISM language} 

\begin{table}[t]
    \caption{Safety (S) and Optimality (O) Constraints}
    \label{safety-constraints}
    \resizebox{\columnwidth}{!}{
    \begin{tabular}{@{}lll@{}}
        \toprule
        \textbf{ID} & \textbf{Constraint} & \textbf{PCTL} \\ \midrule
        $S_1$ & \vtop{\hbox{\strut All agents will be caught with probability $< 0.15$.}} & $P_{< 0.15}(\lozenge captured_{all})$ \\
        $S_2$ & \vtop{\hbox{\strut Agent $1$ will be caught with probability $<0.15$.}} & $P_{< 0.15}(\lozenge captured_1)$ \\
        $S_3$ & \vtop{\hbox{\strut Agent $i$, $i=2,3$ will be caught with probability $< 0.3$.}} & $P_{< 0.3}(\lozenge
        captured_i)$\\
        %\bottomrule
%    \end{tabular}}
%\end{table}

%\begin{table}[t]
%    \caption{Optimality constraints}
    %\label{optimaility-constraints}
%    \resizebox{\columnwidth}{!}{
%    \begin{tabular}{@{}lll@{}}
%        \toprule
%        \textbf{ID} & \textbf{Constraint} & \textbf{PCTL} \\ \midrule
        $O_1$ & \vtop{\hbox{\strut All agents will reach the goal area with probability $\geq  0.8$.}} & $P_{\geq 0.8}(\lozenge goal_{all})$ \\
        $O_2$ & \vtop{\hbox{\strut Agent $1$ will reach the goal with probability $ \geq 0.85$.}} & $P_{\geq 0.85}(\lozenge goal_1)$ \\
        $O_3$ & \vtop{\hbox{\strut Agent $i$, $i=2,3$ will reach the goal with probability $\geq 0.8$.}} & $P_{\geq 0.8}(\lozenge goal_i)$ \\
        $O_5$ & \vtop{\hbox{\strut The expected reward the agents collect collectively is $\geq 7$.}}  & $R_{\geq 7}(\lozenge end_{all})$\\\bottomrule
    \end{tabular}}
\end{table}

%\fb{I would remove all of the following section and discuss ARL in the related work only.} \bg{I agree} \pe{I don't see any problem with that}

%%%%%%%%%%%%%%%%%%%%%%%%%%%%%%%%%%%%%%%%%%%%%%%%%%%%%%%%%%%%%%%%%%%%%%%%

\section{The Verification of (D)RL Behaviors}

In this section we present the main contributions of this work. First, we introduce the motivating example that we use to illustrate the formal machinery as well as for the experimental evaluation in Sec.~\ref{section:evaluation}. Then, we provide an overview of our new \textit{assured multi-agent RL} technique that aims at automatizing and extending to multi-agent settings the ARL method in \cite{AssuranceInReinforcement}. In particular, we introduce abstract Markov game (AMG)
%. \pe{Should we make clearer that the notion of bisimulation and the automated procedure belong to a different section than AMG?}In particular, we 
and define a new notion of {\em bisimulation} that guarantees the preservation of formulae in PCTL between an AMG and its corresponding Markov game. Finally, we provide a procedure to generate such AMGs automatically.

\subsection{Motivating Scenario: the GFC domain} \label{motivatingExample}

\begin{figure}[h]
  \centering
  \includegraphics[width=.9\linewidth]{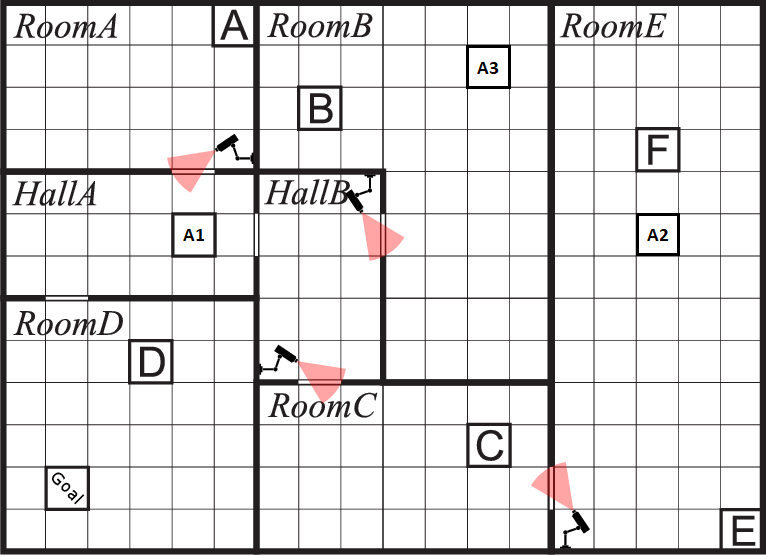}
  \caption{The GFC domain with 3 agents $A_1$, $A_2$, and $A_3$, and 6 flags labelled $A$ to $F$.}
  \label{Flag-Collection-Problem}
  \Description{Figure showing the Guarded Flag-Collection domain from \cite{AssuranceInReinforcement} that we extended with three agents.}
\end{figure}

As motivating scenario we consider the \textit{guarded flag-collection domain} (GFC) from \cite{AssuranceInReinforcement} that we extend with multiple cooperative agents
%. The GFC domain is illustrated in 
(Fig.~\ref{Flag-Collection-Problem}).
The agents' objective is to retrieve as many flags as possible without getting caught by the cameras before reaching the \textit{Goal} position. 
%When an agent gets caught, it is not able to navigate the environment anymore. 
The detection probability of the cameras is given in Table~\ref{detection}. Agents have a different probability of getting caught depending on the action they perform, i.e., hidden, partial or direct. %\fb{we should explain why we have a difference between direct, partial and hidden}. 
%An agent gets a reward of 1 every time it collects a flag and when reaching the goal position. 
An agent's navigation ends when she reaches the goal position or when she gets caught by a camera. An episode terminates when all agents' navigation is over or when the maximum number of step has been reached.  
%\pe{Actually, I think we should not introduce the reward obtained by the agents here. The reason being that at the high level, agents share the same objective and therefore a reward of 1 is obtained for each flag being collected and for each agent that reaches the goal position. We thus express reward constraints on that reward function. However, at the concrete level, agents have independent reward functions. Since we haven't introduce the notion of AMG, maybe it is better to introduces the reward functions later (possibly in the evaluation section). An other solution would be to explain that we in order to express constraints on the objectives, we have a reward function at the abstract level that gives a reward of 1 for each flag collected and for each agent that reaches the goal. We could thus say that for training purpose, the reward functions of the agents are different.}
%\fb{I would leave the reward function for later on.}
%\bg{I would clarify this further, a newcomer to this would ask, can an agent collect more than one flag? you explain when the navigation terminates for one agent but when does an episode terminate? the reward of +1 is when collecting the flag or reaching the goal position?}

This scenario illustrates the need to define agent-specific constraints.
%specific to each robot. 
For instance, 
%we can imagine that 
we may send multiple robots to collect the flags and some robots might be more valuable than others. Consequently, we want to be able to require the robots to have different probabilities to reach the goal position (see, e.g., Table~\ref{safety-constraints}). Further, we observe that the state space grows exponentially with the number of agents: the GFC domain with three agents has $\, \sim 8.2e^{8}$ states versus $\, \sim 1.5e^{4}$ states for the case of a single agent considered in \cite{AssuranceInReinforcement}. This remark strengthens the need for our method to be compatible with deep RL algorithms.
\begin{table}[t]
  \caption{Detection probabilities of the GFC domain cameras.}
  \label{detection}
  \begin{tabular}{@{}cccc@{}}
    \toprule
    & \multicolumn{3}{l}{\textbf{View Detection Probabilities}} \\ \cmidrule(l){2-4} 
    \textbf{Area Transitions} & Direct & Partial & Hidden \\ \midrule
    HallA $\leftrightarrow$ RoomA & 0.18 & 0.12 & 0.06 \\
    HallB $\leftrightarrow$ RoomB & 0.15 & 0.1 & 0.05 \\
    HallB $\leftrightarrow$ RoomC & 0.15 & 0.1 & 0.05 \\
    RoomC $\leftrightarrow$ RoomE & 0.21 & 0.14 & 0.07 \\ \bottomrule
    \end{tabular}
\end{table}

\subsection{Assured Multi-Agent Reinforcement Learning} \label{section:overviewAMARL}

\begin{figure}[h]
  \centering
  \includegraphics[width=\linewidth]{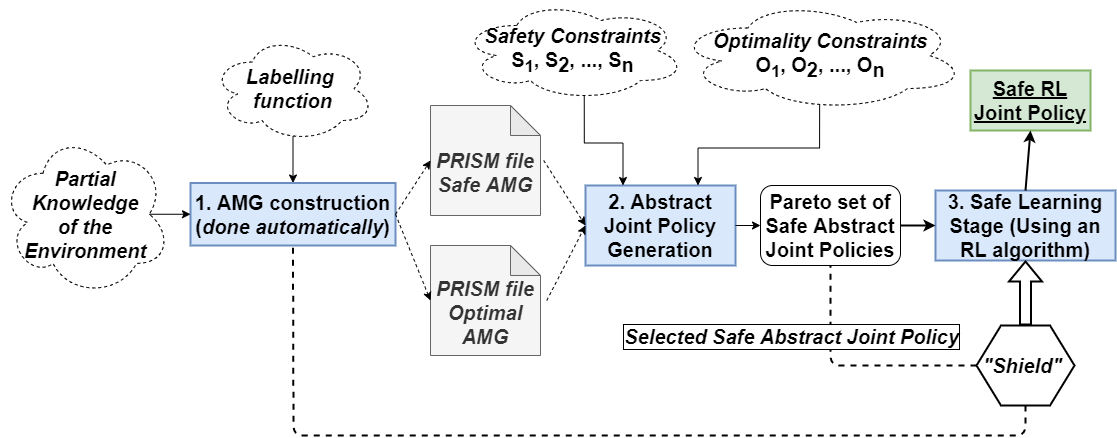}
  \caption{AMARL pipeline, starting with the construction of the abstract MG, then the generation of the abstract joint policies, and terminating with safe learning.}
  \label{Stages-AMARL}
  \Description{Figure showing the process of the AMARL method. Starting with the AMG construction stage, then with the abstract joint-policy generation stage and terminating with the safe learning stage.}
\end{figure}

The main contribution of this paper is the {\em assured multi-agent reinforcement learning} (AMARL) method ensuring formally that teams of autonomous agents satisfy given constraints during the learning process. To this end, we build upon the ARL framework \cite{AssuranceInReinforcement} that we briefly discussed in Sec.~\ref{section:relatedWork}. %\pe{Need to change that once we moved ARL overview to related work.}. 
In this section, we introduce 
%this extension that we call 
%{\em assured multi-agent reinforcement learning} (
AMARL by providing an overview of the method, including the different stages it involves. The AMARL pipeline is depicted in Fig.~\ref{Stages-AMARL}.

\paragraph{AMG construction}

In this stage, the abstract MG corresponding to the given Markov game is generated in an automated manner (see Sec.~\ref{subsec:quotienting}). Indeed, only some domain expertise
%and partial knowledge of the environment are 
is required to define the labelling function. For instance, in the GFC domain (see Fig.~\ref{Flag-Collection-Problem}), it is sufficient to know the lower and upper bounds detection probabilities of the cameras and the layout of the rooms. Note that since the learning stage makes use of algorithms that are not guaranteed to converge to an optimal solution (such as DRL algorithms), we build both an AMG that considers transitions with the higher chance of reaching a "bad state" called {\em safe AMG}, and an AMG that considers transitions with the lower chance of reaching a "bad state" called {\em optimal AMG}. For instance, in the GFC domain
%(see Fig.~\ref{Flag-Collection-Problem}), 
the {\em safe AMG} considers the direct transitions in Table \ref{detection}, whereas the {\em optimal AMG} considers hidden transitions. Nonetheless, the two AMGs are identical but for their transition probabilities and, therefore, have the same abstract policies. %(due to the fact that both AMG have the same successor states and the same number of transitions).

Further, we prove a preservation result on the satisfaction of constraints in PCTL between the AMG and the original MG, by using a bisimulation relation \cite{StrongSimulation, principlesmodelchecking} 
%This stage is explained in detail in 
(see Sec.~\ref{section:abstraction}).
This is a key difference w.r.t.~ ARL, where the abstract model was handcrafted and no preservation guarantees were provided.
%about the preservation of constraints expressed in PCTL. 

%\fb{the following is not clear to me.} Obviously, with full knowledge of the environment, only the optimal AMG is built as we can differentiate similar transitions, and thus able to prevent unsafe actions during the learning stage according to the selected abstract policy.

%Another improvement of AMARL w.r.t.~to ARL consists of the automated generation of the model file corresponding to the AMG, that allows applying model checking techniques on it. \pe{The files generation is easy to implement, and for space reason, we may want to omit to say this.}
%The generation of this file is done straightforwardly according to the states and the transition function of the generated AMG.

\paragraph{Abstract Joint-Policy Generation}

In the second stage of AMARL
%our method is similar to the abstract policy synthesis of ARL. 
we generate arbitrary joint policies in the AMG. 
%Once an abstract joint policy has been generated, 
Then, we use the Storm model checker  \cite{Storm} to verify the relevant constraints in PCTL on the AMGs, suitably restricted according to the joint policy. Safety constraints are verified on the {\em safe AMG}, whereas optimality constraints on the {\em optimal AMG}. (See Table \ref{safety-constraints} for safety and optimality constraints.)
%\fb{I would remove the following. It seems to me a bit of a repetition: The safety constraints do not assume any optimal behaviours from the agents and therefore only express constraints on the upper bound of reaching "bad states" (See Table \ref{safety-constraints}). On the other hand, optimality constraints assume that agents learn an optimal solution and therefore express conditions on the lower bound probability of reaching "good states" and on the amount of reward collected (See Table \ref{optimaility-constraints}). The optimality constraints correspond, therefore, to the same type of constraints that are expressed in ARL \cite{AssuranceInReinforcement}.}
Consequently, the safety constraints allow us to certify that agents always act safely (even during the learning stage, unlike ARL), while the optimality constraints allow us to evaluate the performance of the abstract policies. Finally, as in ARL, we build a Pareto set \cite{liu2014multiobjective} of {\em safe abstract joint policies} that enables us 
%\pe{\sout{to have a trade-off between the different properties of  the safe abstract joint policies}}
to choose a Pareto-optimal safe abstract joint policy according to our preferences.

\paragraph{Safe Learning}

%In the Safe Learning stage, 
Finally, even though we obtained the high-level solution, at the concrete level the selected abstract joint policy can be consistent with several different concrete policies.
%agents still need to learn which action to perform in each state in order to complete the selected abstract joint policy optimally.
We therefore apply deep RL to let agents learn an optimal concrete policy consistent with the selected safe abstract joint policy. Specifically, we use independent deep Q-Learning \cite{IDQN}. Our safe learning stage differs from the corresponding stage of the ARL method on several accounts. Firstly, in order to make our approach compatible with DRL algorithms, we do not remove any action from the action space of the agents. %\pe{Please read the following carefully and let me know if I need to change anything} \fb{It looks fine now.} 
Thus, to make agents learn that some actions are unsafe, we introduce our notion of {\em shield} \cite{safeRLviaShielding} on the environment as depicted in Fig. \ref{Stages-AMARL}. Our shield makes use of the AMG to derive the bisimulation relation (See Def.~\ref{def:weakBisimulationMarkovGames}) between the states of the MG and the states of the AMG, then restrict the behaviours of the agents in the original MG to the selected abstract policy. That is, every time an agent selects an action, before letting the agent interacting with the environment, the shield verifies if the agent's action leads to states allowed by the abstract joint policy and the {\em relaxed bisimulation relation} (see Def.~\ref{relaxBranchingCondition}).
%The shield prevents agents from taking actions that do not belong to the selected abstract policy, nor are incompatible with the conditions expressed by our definition of relaxed stutter bisimulation (see Sec,~\ref{section:relaxed}).
Accordingly, every time the shield block an action, the agent gets a reward of -1 and remains in its current position without interacting with the environment.
%according to the selected abstract policy, and assigns a reward of -1 \fb{and the definition of relaxed stutter bisimulation that we introduce in section 5: an action is dangerous according to the definition of bisimulation? this is not clear}. 
%When an agent tries to perform an unsafe action, this "Shield" prevents it from %taking this action and assign it a reward of -1. 
By doing so, the agent learns that the action is not safe and the shield will no longer be needed at test time when agents follow the learned joint policy.
%, and its optimal policy will never pick this action. 
Differently from ARL, to facilitate the learning process of the agents, the shield
%we introduce 
also recompenses the agents with a reward of 1 every time they complete an option. %\pe{We may want to adapt the presentation of the shield accordingly to the definition of shield provided in Section \ref{section:relaxed}}
%\fb{the rest looks fine to me.}

\subsection{Abstract Markov Games}

Due to the general high dimensionality of the problems solved with (deep) RL algorithms, it is typically not possible to apply model checking techniques directly on the concrete problems. A notion of abstraction is therefore required, and it is then necessary to show the class of formulas that can be verified on the abstract model and preserved in the original one.
In this section we formally define the notion of abstract Markov game used in the AMARL method. However, we first introduce the concepts of {\em options} \cite{options} and {\em termination scheme} \cite{termination} as they are required for the definition of AMG.

%\fb{I think we need to give the definition of option and termination condition before the definition of AMG.}

%\fb{please check the following.}

%\paragraph{Options} 
An {\em option} is  defined as a tuple $o = \langle I_o, \pi_o, \beta_o \rangle$ where $I_o \subseteq S$ is the initialisation set; $\pi_o: S \to A$ is its policy; and $\beta_o \subseteq S$ is the termination condition \cite{options}.
An option can be thought of as a sequence of actions that once initialised follows a policy until it reaches a termination condition. 

%\paragraph{Termination scheme}
When 
%primitive 
actions are replaced with options, given that options, as sequences of actions, might have different duration and termination condition for different agents, thus terminating at different times, it is required to decide when to terminate the joint option.
%\fb{I would just mention $T_{all}$ and remove the other schemes, also from the definition of AMG, to save some space as well.}
%In \cite{termination}, three termination scheme $T_{any}, T_{all}$ and $T_{continue}$ are identified and allow determining when is the next decision epoch of a model.
%
In this paper, we focus on a fully cooperative setting that uses the {\em $T_{all}$ termination scheme} proposed in \cite{termination}, whereby 
%Using the $T_{all}$ termination model means that 
the next joint option is decided as soon as all the options currently being executed have terminated. The reason for this choice is that 
%the $T_{all}$ termination scheme 
it makes the decision problem synchronous and allows the definition of the reward and transition functions at the abstract level. Moreover, the {\em $T_{all}$ termination scheme} is the one that has fewer decision epochs and reduces further the complexity of the problem. Focusing on fully cooperative problems, on the other hand, allows to define the reward function without knowing the particular policy of each agents. For example, in the GFC domain, if two agents try to collect the same flag, it is not necessary to know which agent collects it to define the reward function. 
%In the remainder of this paper, we omit the $T$ parameter in the definition of AMGs as we always use the $T_{all}$ termination scheme.

\begin{definition}[Abstract Markov Game] \label{def:AMG}
     Given a Markov game $M = \langle S, A_1, \ldots , A_n, P, R_1, \ldots ,$ $R_n,$ $\gamma \rangle$, let $O_1, \ldots , O_n$ be a tuple of option spaces over $M$, and $z$ be an abstraction function for $S$. We define the {\em abstract Markov game} corresponding to $M$ as a tuple $\bar{M} = \langle \bar{S}, O_1, \ldots , O_n, \bar{P}, \bar{R_1}, \ldots ,$ $\bar{R_n}, \gamma \rangle$ where:
    \begin{itemize}
        \item $\bar{S}$ is the abstract state space with $z(S) = \bar{S}$.
        \item For every $i \leq n$, $O_i$ is the {\em option space} of agent $i$. %An option can be seen as an abstract action.
        %$O_1, \ldots, O_n$ is the option space of $agent_i$. 
        
        %\pe{I was referring to abstract actions instead of options to put a clear distinction between a sequence of action (options) and an (abstract) action. Thus, allowing us to interpret an AMG as an MG. I think we need to discuss what would be the preferred definition.}
        %\fb{I would just talk about options and mention that these can be seen as "abstract" actions, and therefore an abstract MG is an MG in particular. Otherwise, I am afraid that the distinction option/abstract action could be confusing for some reader.}
%        \item $\bar{A_1}, \ldots , \bar{A_n}$ where $\bar{A_i}$ is the abstract action space of $agent_i$ and correspond to the option space $O_i$ of $agent_i$ in $M$.
 
        \item $\bar{P}$ is the transition function where $\bar{P}^{\vec{\bar{o}}}_{\bar{s}\bar{s'}}$ denotes the probability of going from the abstract state $\bar{s} \in \bar{S}$ to the abstract state $\bar{s'} \in \bar{S}$ with the joint option $\vec{o}= \langle o_1, \ldots , o_n \rangle$: %\pe{This needs to be updated according to the choice we make, either for abstract actions or for options}
        %This joint abstract action $\vec{\rm \bar{a}}$ correspond in $M$ to the joint option $\vec{\rm o}= \langle o_1, \ldots , o_n \rangle$ with $o_i \in O_i$ being the options chosen by all the agents simultaneously.
        $
            \bar{P}^{\vec{\bar{o}}}_{\bar{s}\bar{s'}} = \sum_{s \in S, z(s) = \bar{s}} w_s \sum_{s' \in S, z(s') = \bar{s'}} P(s, \vec{o}, s')
        $
        
            where $w_s$ denotes the weight of state $s$ and represents the degree whereby $s$ contributes to the abstract state $z(s)$ \cite{weightOfState},
            
            \item $\bar{R_1}, \ldots , \bar{R_n}$ is the set of \textit{reward function} where $\bar{R}^{\vec{o}}_{i, \bar{s}\bar{s'}}$ denotes the reward perceived by agent $i$ when performing the 
            %joint abstract action $\vec{\rm \bar{a}}$, corresponding to
joint option $\vec{o} = \langle o_1, \ldots , o_n \rangle$ in $M$, from state $\bar{s} \in \bar{S}$ to state $\bar{s'} \in \bar{S}$. 
%We note $r_{i,t}$ the reward received by agent $i$ at completion of transition $t$: 
        $\bar{R}^{\vec{\rm \bar{o}}}_{i, \bar{s}\bar{s'}} = \sum_{s \in S, z(s) = \bar{s}} w_s R_i(s, \vec{o})$
        \item $\gamma \in [0, 1]$ is the discount factor.
        %\item $n$ is the number of agents.
        %\item \fb{I would just mention that we use $T_{all}$ and remove this point from the definition} $T \in \{T_{any}, T_{all}, T_{continue}\}$ is the termination scheme \cite{termination}.
    \end{itemize}
% $R_i(s, \vec{o})$ is the expected total reward for $agent_i$ until $\vec{o}$ terminates and $P(s, \vec{o}, s')$ is the probability of going to state $s'$ from state $s$ by performing $\vec{o}$.
\end{definition}

By Def.~\ref{def:AMG} AMGs can be considered as a special case of Markov games where an option is considered as an abstract action and we normally omit the bar above the letters in the notation of AMGs.
In Sec.~\ref{subsec:quotienting} we provide an algorithmic way to generate abstract MG so that constraints expressed in wPCTL are preserved, based on the notion of (stutter) bisimulation developed in the following section.

\subsection{Stutter Bisimulations} \label{section:abstraction}

%Due to the general high dimensionality of the problems solved with (deep) RL algorithms, it is typically not possible to apply model checking techniques directly on the concrete problems. A notion of abstraction is therefore required, and it is necessary to show the class of formulas that can be verified on the abstract model and preserved in the original one.

%\pe{Do we use the name stutter bisimulation or stutter bisimulation}
This section is devoted to identifying the conditions under which constraints expressed in PCTL are preserved between an MG and the corresponding abstract MG.
To this end, we first consider \textit{stutter bisimulations}, which are known to preserve the \textit{weak fragment of PCTL} (wPCTL) in stochastic systems 
\cite{StrongSimulation}.
%\fb{does the stutter bisimulation in \cite{StrongSimulation} already preserves wPCTL?} \pe{Yes but the notation they use is a bit different than ours. In fact, they label the actions and not the state.}.
%that when a formula expressed with the   holds in an abstract model that is a stutter bisimulation of a concrete model, this formula also holds in the concrete model. 
In particular, we provide a new notion of stutter bisimulation adapted for Markov games and prove that this new definition preserves the formulas expressed in wPCTL.
%with the \textit{"Weak Fragment of PCTL"} . 
Then, as this direct adaptation of stutter bisimulations to Markov games turns out to be too restrictive for the safe learning stage to be efficient,
%when restricting the actions of the agents to meet both the conditions of the selected abstract policy and of the stutter bisimulation; 
we provide a relaxation of one of the requirements of the stutter bisimulation, 
thus 
%to make the safe learning stage more efficient \fb{can we be more specific, it is just efficient.... efficient.} \pe{The relaxation 
allowing the agents to learn more independently while reducing the number of interventions of the shield. 
%The relaxation also the agents to reduce the number of steps needed to complete the selected abstract joint-policy}. 
%We then show the class of formulas expressed with the \textit{"Weak Fragment of PCTL"} that are preserved by this relaxed definition. 
Finally, we provide a procedure to automatically generate an AMG that is stutter-bisimilar to a given MG.

\subsubsection{Stutter Bisimulation}

Stutter bisimulations have been introduced for probabilistic systems in \cite{StrongSimulation}, where they are referred to as \textit{weak bisimulations}.
Here we extend these bisimulations to Markov games and a multi-agent setting.

We first recall some notations and definitions.
%required for the formal definitions of \textit{Bisimulation} for Markov Games.
\paragraph{Probability distribution.} 
%\cite{AdvancesProbMC} 
For a finite set $X$, a (discrete) {\em probability distribution} on $X$ is defined by a function $\mu: X \rightarrow [0, 1]$ such that $\sum_{s \in S} \mu(s) = 1$. $Dist(X)$ denotes the set of all probability distributions over $X$. 

\paragraph{Notation.} 
%\cite{principlesmodelchecking} 
Let $S$ be a set of states and $\mathcal{E}$ 
%\bg{probably is better to use a different notation here since R was already used for the reward function} \pe{Should we use $E$ for Equivalence relation?} \fb{maybe $\mathcal{E}$ is better, as $E$ is sometimes used for expected values.} 
an {\em equivalence relation} over $S$, i.e., a transitive, reflexive and symmetric binary relation on $S$. For each $s \in S$, $[s]_\mathcal{E} = \{ s' \in S \mid (s, s') \in \mathcal{E} \}$ denotes the equivalence class of state $s$ under $\mathcal{E}$. Then, $S \backslash \mathcal{E} = \{[s]_\mathcal{E} \mid s \in S\}$ is the {\em quotient space} of $S$ under $\mathcal{E}$.
%is the set consisting of all $R$-equivalence classes. 
For a relation $\mathcal{E}$, if ($s_1, s_2) \in \mathcal{E}$, we often write $s_1 \mathcal{E} s_2$. Let $M$
%= \langle S, A_{1}, \ldots , A_{n}, P, R_{1}, \ldots , R_{n}, \gamma, n \rangle$ 
be a Markov game, $P(s, \vec{a}, \cdot)$ denotes the discrete probability distribution on $S$ to select the next states, given the current state $s$ and joint action $\vec{a}$.
\begin{definition}[$\mathcal{E}$-equivalence \cite{StrongSimulation}]
    Let $\mathcal{E}$ be an equivalence relation over set $S$ and let $\mu_1, \mu_2 \in Dist(S)$ be two probability distribution. We say that $\mu_1$ and $\mu_2$ are \textit{$\mathcal{E}$-equivalent}, denoted $\mu_1 \equiv_\mathcal{E} \mu_2$, iff $\mu_1([s]_\mathcal{E}) = \mu_2([s]_\mathcal{E})$, for all $[s]_\mathcal{E} \in S \backslash \mathcal{E}$, where $\mu([s]_\mathcal{E}) = \sum_{s' \in [s]_\mathcal{E}} \mu(s')$. That is, $\mu_1$ and $\mu_2$ are $\mathcal{E}$-equivalent if they assign the same probability weight to every $\mathcal{E}$-equivalence class.
\end{definition}

Having recalled the required notations,
%and the notion of an equivalence relation from \cite{StrongSimulation}, 
we now introduce our novel definition of \textit{stutter bisimulation} specifically for Makov games. 
%\fb{if novel, what is the difference wrt~ \cite{StrongSimulation}.} \pe{See above. But also I provided a short explanation of the changes after the definition. Should we change this?}
%
%\fb{rather than two MGs, I'd given the following definition directly for an MG and the corresponding AMG.}
\begin{definition}[Stutter Bisimulation] \label{def:weakBisimulationMarkovGames}
    Let $M_i = \langle S_i, A_{1, i}, \ldots,$ $A_{n, i}, \allowbreak P_i, R_{1, i}, \ldots , R_{n, i}, \gamma_i \rangle$, for $i = 1, 2$, be two Markov games %\fb{let's just assume that our MG do not have terminal states and eliminate the following: without terminal states} 
    over $AP$ with labelling functions  $L_i: S_i \rightarrow 2^{AP}$ for $M_1$ and $M_2$. A \emph{(stutter) bisimulation} for $(M_1, M_2)$ is an equivalence relation $\mathcal{E}$ over $S_1 \cup S_2$ such that
    \begin{enumerate}
        \item for every $s_1 \in S_1$, there exists $s_2 \in S_2$ such that $s_1 \mathcal{E} s_2$. %\fb{initial states are not defined in $M_1$ and $M_2$, we can just replace $I_i$ with $S_i$.}
        \item for every $s_2 \in S_2$, there exists $s_1 \in S_1$ such that $s_1 \mathcal{E} s_2$.        
        \item for all $(s_1, s_2) \in \mathcal{E}$ it holds that:
        \begin{enumerate}
            \item $L_1(s_1) = L_2(s_2)$,
            \item for every joint action $\vec{a}_1$ available from state $s_1$, there exist a joint option $\vec{o}_2$ available from $s_2$ such that $P_1(s_1, \vec{a}_1, \cdot) \equiv_\mathcal{E} P_2(s_2, \vec{o}_2, \cdot)$ and $\vec{o}_2$ respect the branching condition \cite{branchingCondition}: 
            %\pe{Please have a look to the following and let me know if there is anything unclear. Note that we use joint option to match a joint action. It is an important attribute of the definition.}
            \begin{itemize} %\fb{It looks fine to me.}
                \item for every path $w$ consistent with 
                %that can occur when following 
                $\vec{o}_2$ 
                %\fb{what does it mean this notation? how can a path belong to an option?} \pe{I reformulated it in words, is it clearer?} 
                and every state $s$ occurring in $w$, either $s_1 \mathcal{E} s$ and each state $s'$ that precedes $s$ in $w$ satisfies $s_1 \mathcal{E} s'$, or for every $s'_1 \in S_1$, $s'_1 \mathcal{E}$ $last(w)$ implies $s'_1 \mathcal{E} s$.
            \end{itemize}
            
            \item for every joint action $\vec{a}_2$ available from state $s_2$, there exist a joint option $\vec{o}_1$ available from $s_1$ such that $P_2(s_2, \vec{a}_2, \cdot) \equiv_\mathcal{E}  P_1(s_1, \vec{o}_1, \cdot)$ and $\vec{o}_1$ respect the branching condition \cite{branchingCondition}:
            \begin{itemize}
                \item for every path $w$ consistent with
                %that can occur when following 
                $\vec{o}_1$ and every state $s$ occurring in $w$, either $s_2 \mathcal{E} s$ and each state $s'$ that precedes $s$ in $w$ satisfies $s_2 \mathcal{E} s'$, or for every $s'_2 \in S_2$, $s'_2 \mathcal{E}$ $last(w)$ implies $s'_2 \mathcal{E} s$.
            \end{itemize}
        \end{enumerate}
    \end{enumerate}
    where 
    %$I_i \subseteq S_i$ denotes the set of initial states of $M_i$, and 
    $last(w)$ denotes the last state of path $w$. 
    
    We write $M_1 \simeq M_2$ whenever there exist a bisimulation $\mathcal{E}$ for $(M_1, M_2)$, and for every $(s_1, s_2) \in \mathcal{E}$ we write $s_1 \simeq s_2$.
\end{definition}

%\fb{where do we say that an MG and the AMG are stutter bisimilar?}
%\pe{As we said before that an AMG could be seen as an MG, I assumed it was enough to introduce the relation with two MGs. Do you think we should make it more explicit?} \fb{I've added a few lines at the end of Sec.~3.4. But it would be good to recall this point where necessary.}

\begin{figure} 
    \centering
    \resizebox{.66 \linewidth}{!}{%
    \begin{tikzpicture}
        \node[state, initial] (v0) {$v_{0}, \{a\}$};
        \node[state, right of=v0] (v1) {$v_{1}, \{a\}$};
        \node[state, right of=v1] (v2) {$v_{2}, \{a\}$};
        \node[state, right of=v2, below=5pt] (v3) {$v_{3}, \{b\}$};
        \node[state, right of=v2, above=5pt] (v4) {$v_{4}, \{c\}$};
        
        \draw 
              (v0) edge[bend right, below] node{$\overrightarrow{x}$, 1} (v1)
              (v1) edge[bend right, above] node{$\overrightarrow{x}$, 1} (v0)
              (v1) edge[left, above] node{$\overrightarrow{y}$, 1} (v2)
              (v2) edge[left, below] node{$\overrightarrow{x}$, 0.7} (v3)
              (v2) edge[left, above] node{$\overrightarrow{x}$, 0.3} (v4)
              (v2) edge[loop above] node{$\overrightarrow{y}$, 1} (v2)
              (v3) edge[loop below] node{$\overrightarrow{x}$, 1} (v3)
              (v4) edge[loop above] node{$\overrightarrow{x}$, 1} (v4);
    \end{tikzpicture}}
 \hspace{.1cm}   
    \resizebox{.31\linewidth}{!}{%
    \begin{tikzpicture}
        \node[state, initial] (s0) {$s_{0}, \{a\}$};
        \node[state, right of=s0, below=5pt] (s1) {$s_{1}, \{b\}$};
        \node[state, right of=s0, above=5pt] (s2) {$s_{2}, \{c\}$};
        
        \draw 
              (s0) edge[right, above] node{$\overrightarrow{x}$, 0.3} (s2)
              (s0) edge[right, below] node{$\overrightarrow{x}$, 0.7} (s1)
              (s0) edge[loop above] node{$\overrightarrow{y}$, 1} (s0)
              (s2) edge[loop above] node{$\overrightarrow{x}$, 1} (s2)
              (s1) edge[loop below] node{$\overrightarrow{x}$, 1} (s1);
    \end{tikzpicture}}
    
    \caption{The MG $M_1$ on the left and its stutter-bisimilar AMG $M_2$ on the right.}
    \Description{Figure showing the stutter bisimulation relation on two MGs.}
    \label{fig:AMG_Bisimilar}
\end{figure}
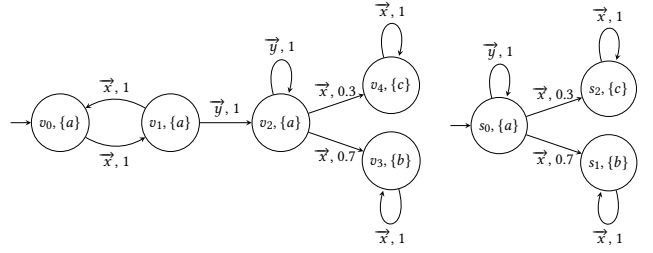

Conditions (1) and (2) in Def.~\ref{def:weakBisimulationMarkovGames} state that every state of $M_1$ is in relation with some state of $M_2$ and vice versa. Condition (3.a) requires that bisimilar states are equally labelled. Condition (3.b) says that every outgoing transition $P_1(s_1, \vec{a}_1, \cdot)$ must be $\mathcal{E}$-equivalent to some outgoing transition $P_2(s_2, \vec{o}_2, \cdot)$. 
%All the possible paths of joint-option $\vec{\rm o_2}$ can be seen as a tree where the root is the initial state of $\vec{\rm o_2}$.\fb{in what sense?} \pe{Is it clearer? What I mean by a tree is that it represents all the possible paths that can occur when following the policy of $\vec{\rm o_2}$. Thus, each node represent a state and its children are the possible successors of this state when following the policy of $\vec{\rm o_2}$}. 
The branching conditions says that for every path $w$ that can occur when following $\vec{o}_2$, all the states in $w$ that appear before the change of equivalence class are related to state $s_1$, whereas all the states that appear thereafter
%the change of equivalence class in the path 
are related to the same state $s'_1 \in S_1$.
%The branching condition states that all the states of a path $w$ that occur between the root and the change of equivalence class in the tree are related to state $s_1$ and that all the states that occur after this change are related to some state $s'_1 \in S_1$. 
Condition (3.c) is the symmetric counterpart to (3.b). 
%
%\fb{I think it would be good to have a figure depicting a relation of stutter bisimulation between and MG and the corresponding AMG.}
%
%\pe{
As an example, Fig.~\ref{fig:AMG_Bisimilar} depicts an MG $M_1$ and its corresponding AMG $M_2$, which is stutter-bisimilar according to the equivalence relation that induces the following set of equivalence classes $\{ \{v_{0}, v_{1}, v_{2}, s_{0}\},$ $\{v_{3}, s_{1}\}, \{v_{4}, s_{2}\}\}$. %\fb{I've resized the figure. Please check that you're happy with that.}

%\pe{Please also check the paragraph above. I think you made the change from joint option to joint action, let me know if there was a specific reason or if anything is unclear.} \fb{all looks fine to me now.}

Our definition of stutter bisimulation can be seen as a unified version of the stutter bisimulation for deterministic systems in \cite{principlesmodelchecking} and the weak bisimulations for probabilistic systems in \cite{StrongSimulation}. Indeed, our definition takes into account the state labellings as in \cite{principlesmodelchecking} and supports the probabilistic case as in \cite{StrongSimulation}. However, a notable difference here is the use of options which follow memoryless policies, in contrast with the notions in \cite{principlesmodelchecking, StrongSimulation} that use history-based policies. %\fb{provide a reference for other definitions?}.
%taking history into account. 
Further, conditions (3.b) and (3.c) in 
%the last difference is that 
%our notion of bisimulation 
Def.~\ref{def:weakBisimulationMarkovGames} guarantee that the equivalence relation $\mathcal{E}$ is {\em divergence-sensitive}, i.e., the following lemma holds.
%if there is an infinite path starting in $s$ that only visits states in $[s]_R$ for every $s'$ such that $(s, s') \in R$, there also exist an infinite path starting in $s'$ that only visits states in $[s]_R$. The satisfaction of this property is assured by condition (3.b) and (3.c) in Def.~\ref{def:weakBisimulationMarkovGames}. 
%
\begin{lemma} \label{lemma:divergence-sensitive}
Let $M_i = \langle S_i, A_{1, i}, \ldots,$ $A_{n, i}, \allowbreak P_i, R_{1, i}, \ldots , R_{n, i}, \gamma_i \rangle$, for $i = 1, 2$, be two bisimilar Markov games, with bisimilar states $s_1 \in S_1$ and $s_2 \in S_2$.
%    By conditions (3.b) and (3.c) of Def.~\ref{def:weakBisimulationMarkovGames}, for any state $s \in S$, if there exist an 
For every infinite path $w_1$ from $s_1$ that always remains in $[s_1]_\mathcal{E}$, 
%then for any $s' \in [s]_R$ 
there exists an infinite path $w_2$ from $s_2$ that always remains in $[s_2]_\mathcal{E}$ (and viceversa).
\end{lemma}
%\fb{the following can be restated as a lemma before the main theorem.} \pe{See lemma \ref{lemma:divergence-sensitive}}
%It is easy to see that because conditions (3.b) and (3.c) require that every transition available from a state $s \in [s]_R$ can be mimicked by an option from any state $s' \in [s]_R$, then if there exists an infinite path from $s$ that always remain in $[s]_R$, there will also exist an infinite path from $s'$ that always remain in $[s]_R$. 

%Finally, note that our notion of \textit{bisimulation} is more restrictive than both definitions in \cite{StrongSimulation} and \cite{principlesmodelchecking} in the sense that our definition does not only express conditions on transitions that leave the equivalence class but on all the transitions.
%of a state can be mimicked by an option of any equivalent state sequence of transitions of an equivalent state even if the transition remains in same the equivalence class. 
%\fb{the above is not really clear. Can it be restated?} \pe{Is it clearer now? Also, maybe we can remove it if we need space as I don't think it is a major remark.}

Note that we provide in Appendix \ref{appendixA} all the relevant proofs of the Lemmas and Theorems we introduce in this paper. We now state the main theoretical result in this section about the preservation of wPCTL under stutter bisimulations.

\begin{theorem} \label{weakBisimulationTheoremMG}
    Let $M_1$ and $M_2$
    %= \langle S_i, A_{1, i}, \ldots , A_{n, i}, P_i, R_{1, i}, \ldots , R_{n, i}, \gamma_i \rangle$, 
    be two Markov games 
    %without terminal states 
    over $AP$ with labelling functions $L_i: S_i \rightarrow 2^{AP}$ for $M_1$ and $M_2$. 
    For every weak PCTL formula $\Phi$, we have that if $M_1 \simeq M_2$, then $M_1 \models \Phi \iff M_2 \models \Phi$.
\end{theorem}

%\begin{proof}
%    The proof of Theorem \ref{weakBisimulationTheoremMG} is done by induction over the structure of $\Phi$
%    \bg{the proof environment should include the whole proof or have a reference to it if iy is to appear in the supplemental material}
%\end{proof}

By Theorem~\ref{weakBisimulationTheoremMG} two bisimilar Markov games 
satisfy the same  formulae in wPCTL. This is a key result that allows us to guarantee that, since we can build AMG that are bisimilar to the concrete MG,
 then all  properties expressed in the weak fragment of PCTL that hold in the AMG, also hold in the corresponding MG. However, it is important to note that the branching condition expressed in items (3.b) and (3.c) of Def.~\ref{def:weakBisimulationMarkovGames} 
%during the execution of a joint-option, 
%does not allow agents to reach states that are neither bisimilar with a terminal state of this joint-option nor to its initial state. Consequently, this condition
 forces agents to coordinate to perform a joint action that transition from a state bisimilar to the initial state of the joint option to a state bisimilar to a terminal state of it, thus preventing from terminating their options independently. As we assume that 
 %in an RL environment an 
 agents can perform 
 an idle action that does nothing,
 %i.e., the agent does not interact with the environment; 
 this condition does not have any impact theoretically. On the other hand, using the shield 
 %of our method 
 to guarantee the satisfaction of the branching condition would considerably complicate the learning process and deteriorate the quality of the learned solution. For this reason, in the next section, we introduce a relaxed version of the branching condition that allows agents to 
 %interact with the environment and 
 terminate their options independently, thus improving the quality of the learned solution and facilitating the learning stage by reducing the number of interventions of the shield,
% \fb{in what ways the number of interventions by the shield is reduced?} \pe{Since the relaxed version is less restrictive, 
as the agents are normally less likely to violate the constraints for the relaxed version.

\subsubsection{Relaxing the Branching Condition} \label{section:relaxed}

In a multi-agent RL environment 
%as agents interact independently, 
it is key to verify the behaviour of each agent independently, e.g., we want to be able to ensure that a specific agent is not reaching a dangerous state. 
%To this end, we can specify atomic propositions that are specific to an agent, and that does not impact the other agents. (e.g. in our GFC domain, the AP that is specific to each agent is the position of the agent.) 
However, some atomic propositions expressing overall goals the agents want to achieve, will be shared by all agents.
%and are required to represent the complete problem we want to solve. 
E.g., in the GFC domain, the flags that have been collected are represented as atoms that are shared amongst all agents as it is a common goal.
%\fb{what shall we do with the information above?}
%According to this discussion, we introduce the following notations:
%
%\paragraph{Notations.} 
Following this idea, we denote as $AP_{i}$ the set of atoms whose truth depends only on the local state of agent $i$, such as its position, and $AP_{all}$ the set of atoms whose truth depends on the whole global state of the system. %Formally, the set $AP_{agent_i}$ is the set of atomic propositions that can only change with respect to the actions taken by $Agent_i$. On the other hand, the set $AP_{all}$ is the set of atomic propositions that can be affected by the actions of any agents. It is therefore straightforward to define these sets of $AP$ beforehand, and in the remainder of this project, we assume that the labelling function allows distinguishing these sets of atoms.
%
%Once we defined the various $AP_{i}$ and $AP_{all}$, 
Then, we relax the branching conditions (3.b) and (3.c) in Def.~\ref{def:weakBisimulationMarkovGames}. Note that this relaxed branching condition is only used during the learning stage to define the shield %\fb{in what sense define? I think we should define the shield earlier at this point. This is to be solved as discussed at the meeting.} 
and not during the generation of the abstract model. Thus, the generated abstract model is still stutter bisimilar to the concrete model.
\begin{definition}[Relaxed (Stutter) Bisimulation] \label{relaxBranchingCondition}
    Let $M_i = \langle S_i, A_{1, i}, \allowbreak \ldots , A_{n, i}, P_i, R_{1, i}, \ldots , R_{n, i}, \gamma_i \rangle$, $i = \{1, 2\}$ be two Markov Games
    %without terminal states 
    over $AP$ with   labelling functions $L_i: S_i \rightarrow 2^{AP}$ for $M_1$ and $M_2$. A \emph{relaxed (stutter) bisimulation} for $(M_1, M_2)$ is an equivalence relation $\mathcal{E}$ over $S_1 \cup S_2$ such that 
    %and let $R$ be a bisimulation for $M_1$ and $M_2$. $R$ becomes a \textbf{relaxed bisimulation} for $M_1$ and $M_2$ if we replace the branching condition of 
    all conditions in Def.~\ref{def:weakBisimulationMarkovGames} hold for the following relaxed branching condition:
    \begin{itemize}
        \item for every path $w$ consistent with $\vec{o}_k$ and every state $s$ occurring in $w$, 
        \begin{enumerate}
             \item for each agent $i$, 
             %$i=1,2,3,\ldots n$, 
             either $L(s) \cap AP_{i} = L(s_m) \cap AP_{i}$ and each state $s'$ that precedes $s$ in $w$ satisfies $L(s') \cap AP_{i} = L(s_m) \cap AP_{i}$, or for every $s'_m \in S_m$, if $L(s'_m) \cap AP_{i} = L(last(w)) \cap AP_{i}$ then $L(s'_m) \cap AP_{i} = L(s) \cap AP_{i}$
            \item 
            %for every atom $q \in AP_{all}$, 
            either $L(s) \cap AP_{all} = L(s_m) \cap  AP_{all}$ and each state $s'$ that precedes $s$ in $w$ satisfies $L(s') \cap  AP_{all} = L(s_m) \cap  AP_{all}$, or for every $s'_m \in S_m$, if $L(s'_m) \cap  AP_{all} = L(last(w)) \cap  AP_{all}$ then $L(s'_m) \cap  AP_{all} = L(s) \cap  AP_{all}$.
        \end{enumerate}
    \end{itemize}
   where $last(w)$ denotes the last state of path $w$, $k, m \in \{1, 2\}, k \neq m$. 
   
   We write $M_1 \simeq_r M_2$ whenever there exist a relaxed bisimulation $R$ for $(M_1, M_2)$,
   %where the branching condition is replaced by the above conditions,
   and for $(s_1, s_2) \in \mathcal{E}$, we write $s_1 \simeq_r s_2$.
\end{definition}

Intuitively, the difference between the relaxed condition in Def.~\ref{relaxBranchingCondition} and items (3.b) and (3.c) in Def.~\ref{def:weakBisimulationMarkovGames} is that,
%the former, 
instead of ensuring that between the initial and the last state of a path all atoms either change simultaneously or do not change, only ensures that the atoms in each $AP_{i}$, 
%with i=1,2\ldots ,n, 
either change at the same time or do not change, 
%Additionally, the relaxed branching condition says that 
with the atoms belonging to set $AP_{all}$ considered independently.

%According to this relaxation, 
We now state the following theorem,
%Theorem \ref{AgentAPrelaxedStutterBisimulationTheoremMG} concerning the wPCTL properties that are preserved by the relaxed bisimulation, that is, 
which adapts Theorem~\ref{weakBisimulationTheoremMG} to formulae
%properties that only 
containing atoms of a single agent, as well as
%Then, we present Theorem \ref{RelaxedWeakBisimulationTheoremMG} regarding the wPCTL properties that are preserved by the relaxed bisimulation 
formulae with no restriction on the atoms.
%that these properties contain.
%
\begin{theorem} \label{AgentAPrelaxedStutterBisimulationTheoremMG}
    Let $M_1$ and $M_2$
    %= \langle S_i, A_{1, i}, \ldots , A_{n, i}, P_i, R_{1, i}, \ldots , R_{n, i}, \gamma_i, n \rangle$, $i = 1, 2$ 
    be two Markov games 
    %without terminal states 
    over $AP$ with labelling functions $L_i: S_i \rightarrow 2^{AP}$ for $M_1$ and $M_2$.
    \begin{enumerate}
    \item 
    For every wPCTL formula $\Phi$ that only contains atoms in $AP_i$
    %of a single agent, i.e. that only contains AP of a set $AP_{agent_i}$, 
    we have that if $M_1 \simeq_r M_2$, then $M_1 \models \Phi \iff M_2 \models \Phi$.
%\end{theorem}
%
%\begin{theorem} \label{RelaxedWeakBisimulationTheoremMG}
%    Let $M_1$ and $M_2$
    %= \langle S_i, A_{1, i}, \ldots , A_{n, i}, P_i, R_{1, i}, \ldots , R_{n, i}, \gamma_i, n \rangle$, $i = 1, 2$ 
    %be two Markov games without terminal states over $AP$ and let $L_i: S_i \rightarrow 2^{AP}$ be two labelling function for $M_1$ and $M_2$ respectively. 
    \item For every wPCTL formula $\Phi$ where path formulae are restricted to $(true$ $U$ $\Phi)$ and $(false$ $U$ $\Phi)$, i.e., $\lozenge \Phi$ and $\Phi$, we have that if $M_1 \simeq_r M_2$, then $M_1 \models \Phi \iff M_2 \models \Phi$. Moreover, assuming that $\Phi$ only contains one atom and no restriction on path formulae, if $M_1 \simeq_r M_2$, then $M_1 \models \Phi \iff M_2 \models \Phi$.
    \end{enumerate}
\end{theorem}
%
%\begin{proof}
%    Again, the proofs of Theorem \ref{AgentAPrelaxedStutterBisimulationTheoremMG} and Theorem \ref{RelaxedWeakBisimulationTheoremMG} are done by induction over the structure of $\Phi$. \bg{again provide the reference that this is in the supplemental material}
%\end{proof}

We, therefore, defined a relaxation for the branching condition of bisimulation that allows to construct more easily the shield in the safe learning stage. However, it is worth to mention that even though the shield can enable agents to perform actions that lead to a state that is neither bisimilar to the initial state nor to a terminal state of the joint option in the sense of Def.~\ref{def:weakBisimulationMarkovGames}, still it has to verify that after performing an action the global probability of reaching the target equivalence classes remains the same. 
%This last condition is not a problem in practice as we compute the probabilities of reaching each equivalence class from a state when building the abstract model. \pe{It might be needed to rewrite the previous paragraph is a clearer way, do you think it is clear enough?}

%\fb{I think we need a section dedicated to presenting the shield. Now we do bits here and there, but it might be confusing. Again, this is to be solved as discussed at the meeting.}
%\pe{Please also check the following} \fb{Ok.}

Thus, as mentioned in Sec,~\ref{section:overviewAMARL}, our shield is built according to the relaxed version of the branching condition and by Theorem \ref{AgentAPrelaxedStutterBisimulationTheoremMG}, it ensures the 
%allows guaranteeing the 
preservation of some wPCTL formulae both during the learning stage and by the learned solution.

%Thus, our notion of {\em shield} as depicted in Fig.~\ref{section:overviewAMARL} uses the built AMG to obtain the equivalence relation between the states of the MG and the states of the AMG to restrict the behaviours of the agents to the selected abstract policy. Our shield ensure that the actions taken by the agents respect the conditions of the relaxed stutter bisimulation according to the selectd abstract policy. It, therefore, assign a reward of -1 when an agent tries to perform an action that violates these conditions and prevent the agent to interact with the environment. In addition, every time an acton complete its option, the shield assign a reward of 1 to the agent to facilitate the learning process. %\pe{I tried here to give a more formal definition of the shield. However, the part about the reward assigned by the shield are also explained in the overview of AMARL. Should we only present the shield briefly in the overview and keep the explanation above? Or should we do something different?} \fb{To be solved as discussed at the meeting.}

\subsection{Building Bisimilar Abstract Markov Games} \label{subsec:quotienting}

In this section we provide an algorithm to construct the abstract MG corresponding to a given Markov game, so that the AMG is stutter bisimilar to the original MG. %and therefore, by Theorem~\ref{weakBisimulationTheoremMG}, satisfies the same formulae in wPCTL. 
%\pe{Is it really usefull to say that they satisfie the same formulae in wPCTL given that in the end the shield use the relaxed bisimulation?}
%\fb{I've removed that bit.}

%\pe{Should we mention the theorem of stutter bisimulation? Given that we use the relaxed version to define the shield, we use Theorem \ref{AgentAPrelaxedStutterBisimulationTheoremMG} }

%\fb{we might mention that the abstraction function $z$ introduced in the definition of AMG, maps each state $s$ to its equivalence class according to the interpretation of atoms.} \pe{I think this is actually represented by $\Pi_{AP}$ in Algorithm \ref{algo:quotienStutterBisimulation}.} \fb{I think we need to say this explicitly.}

As the starting point, we consider the algorithm given in \cite{principlesmodelchecking} to compute the quotient transition system under stutter bisimulation for deterministic systems, which  is based on the partition refinement technique
%Partition refinement algorithm is a standard method to generate abstract models, and this technique is also suggested in 
\cite{AdvancesProbMC}. We, therefore, adapt this algorithm 
%e quotienting algorithm for deterministic systems in \cite{principlesmodelchecking} 
to compute AMGs under stutter bisimulation. 
%In what follows, we adjust the definitions of the quotienting algorithm for deterministic systems in \cite{principlesmodelchecking} to make them work with MGs. 
Throughout this section, let $M = \langle S, A_{1}, \ldots , A_{n}, P, R_{1},$ $\ldots , R_{n} \gamma \rangle$ be a Markov game 
%without terminal states 
over AP with labelling function $L: S \rightarrow 2^{AP}$. Moreover, the abstraction function $z$ we introduced in Def. \ref{def:AMG} maps each state $s$ to its equivalence class according to the interpretation of atoms.
\begin{definition}[Splitter]
    Let $\Pi$ be a partition of $S$ and let $B \in \Pi$, we have that:
    \begin{enumerate}
        \item A transition $P(s, \vec{a}, \cdot)$, for $s \in B$, $\vec{a} = \langle a_1, \ldots , a_n \rangle$, $a_i \in A_i$ is a $\Pi$-$splitter$ for $B$ iff there exists a state $s' \in B$ that does not have any joint option respecting the branching condition 
        %\cite{branchingCondition} 
        that matches $P(s, \vec{a}, C)$ for every $C \in \Pi$.
        \item $\Pi$ is $B$-$stable$ if there is no $\Pi$-$splitter$ for $B \in \Pi$.
        \item $\Pi$ is $stable$ if $\Pi$ is $B$-$stable$ for all blocks $B \in \Pi$.
    \end{enumerate}
\end{definition}

In words, $P(s, \vec{a}, \cdot)$ is a $\Pi$-$splitter$ for $B$ if it violates conditions (3.b) and (3.c) in Def.~\ref{def:weakBisimulationMarkovGames}, i.e.,  there exist a state $s' \in B$ from where there exist no joint option respecting the branching condition that can mimic the transition $P(s, \vec{a}, \cdot)$. 
%We say that a block $B \in \Pi$ is stable if there does not exist any $\Pi$-$splitter$ for this block $B$, i.e., there does not exist any transition for this block that violates conditions (3.b) and (3.c) of Definition~\ref{def:weakBisimulationMarkovGames}. Finally, we have that if all block $B \in \Pi$ are stable, $\Pi$ is also stable. 
Given that the initial partition obtained with the abstraction function $z$ groups the states in blocks that share the same labelling of atoms, it is easy to see that if $\Pi$ is stable, then $\Pi$ is a stutter bisimulation.
%for $S$.

Once introduced the notion of splitter, 
%it remains to 
we show how to split a block $B \in \Pi$ according to a transition $P(s, \vec{a}, \cdot)$ that was identified as a $\Pi$-$splitter$ of $B$. Thus, we define the function $Split(B, P(s, \vec{a}, \cdot))$ = $\{ B \cap Sat_B(P(s, \vec{a}, \cdot)), B \backslash Sat_B(P(s, \vec{a}, \cdot)) \}$, where $Sat_B(P(s, \vec{a}, \cdot))$ denotes the set of states in $B$ that have a joint option respecting the branching condition that can mimic transition $P(s, \vec{a}, \cdot)$. 
%From this definition, 
We can now define the function $Refine(\Pi, B, \delta)$ that refines a partition $\Pi$.
\begin{definition}[Refinement] \label{ref:refinement}
    Let $\Pi$ be a partition for state space $S$, $B \in \Pi$, and $P(s, \vec{a}, \cdot)$ a $\Pi$-$splitter$ of $B$.
    Then,
    \begin{center}
        $Refine(\Pi, B, P(s, \vec{a}, \cdot))$ = $Split(B, P(s, \vec{a}, \cdot)) \cup (\Pi \backslash B)$
    \end{center}
\end{definition}

By using Def.~\ref{ref:refinement}, we present in Algorithm \ref{algo:quotienStutterBisimulation}  the refinement procedure for quotienting Markov games according to our stutter bisimulation, which is an adaptation of 
%for MGs which is an adapted version of 
the algorithm for 
stutter bisimulation quotienting in \cite{principlesmodelchecking}. 
%Given that the number of iterations of the while-loop is bounded by the number of states, it is required to have an efficient implementation of the $Refine$ function and an efficient way to find the $\Pi$-$splitters$. However, the above definitions and algorithms provide the theoretical framework to generate the stutter bisimilar AMG of a given MG automatically.

%\fb{it is not clear to me why the following is the case or why it should be sufficient anyway.}
Note that we are only interested in the generation of stutter bisimulation and not in relaxed stutter bisimulation as the definition of the relaxed stutter bisimulation is based on a relation that is a stutter bisimulation. 
%\pe{
The relaxed stutter bisimulation is less strict than the stutter bisimulation, thus if we have a relation that is a stutter bisimulation, this relation also fullfil the conditions of the relaxed stutter bisimulation. The only difference is that the relaxed stutter bisimulation allows to reach the next abstract state by going through some abstract states that are not in relation with the initial one nor the final one. From an AMG that is a stutter bisimulation, it is easy to build a shield that restrict the actions of the agents according to the relaxed version of the branching condition.
%}

\begin{algorithm}[h]
    \SetAlgoLined
    \SetKwInOut{Input}{input}\SetKwInOut{Output}{output}
    \Input{An MG $M$ without terminal states over $AP$ and its corresponding labeling function $L$}
    \Output{Stutter bisimulation quotient space $S/\simeq$}
        $\Pi := \Pi_{AP}$ \Comment*[r]{$\Pi_{AP}$ is the initial partition that aggregate states with the same atoms. An algorithm to compute $\Pi_{AP}$ is provided in \cite{principlesmodelchecking}.}
        \While{$\exists B \in \Pi, \exists P(s, a, \cdot), s \in B.$ such that $P(s, a, \cdot)$ is a $\Pi$-$splitter$ for $B$}{
            choose such $B \in \Pi, P(s, a, \cdot), s \in B$ \\
            $\Pi := Refine(\Pi, B, P(s, a, \cdot))$ 
        }
        \Return $\Pi$
     \caption{Algorithm to compute the stutter bisimulation quotient}
     \label{algo:quotienStutterBisimulation}
\end{algorithm}

% \begin{itemize}
%     \item Definition of Stutter Bisimulation and definition of the Weak Fragment of PCTL (We may not have the space to provide all the formal proofs)
%     \item Explain why the original Stutter bisimulation is too restrictive for the learning stage
%     \item Definition of the relaxation and the fragments that it preserves
%     \item The theoretical framework to automatically generate stutter bisimilar AMG.
% \end{itemize}

% \begin{itemize}
%     \item Figure
%     \item Short explanations and comparison of the stages with ARL
% \end{itemize}

%%%%%%%%%%%%%%%%%%%%%%%%%%%%%%%%%%%%%%%%%%%%%%%%%%%%%%%%%%%%%%%%%%%%%%%%

\section{Experimental Evaluation}

\label{section:evaluation}

We evaluate empirically the performance of our method on the GFC domain with three agents presented in Sec.~\ref{motivatingExample}, against the safety and optimality constraints in Table~\ref{safety-constraints}. 
All our experiments are run using an Nvidia Tesla (12GB RAM) - 24-core/48 thread Intel Xeon CPU with 256GB RAM. The DRL algorithm we use is IDQL \cite{IDQN} with Double Deep Q-Learning \cite{doubledeepqlearning} and the Dueling Network Architecture \cite{dueling}. We refer to Appendix \ref{appendixB} for the details on hyperparameters. %\pe{I assume we have to provide a reference to this supplementary material.}. 
Each final policy evaluation is repeated $1e^{4}$ times.

For our experiments, the \textit{fully cooperative} reward function of the AMGs is defined as follows: a reward of 1 is obtained for each flag collected and for each agent that reaches the goal position of the environment. It is important to realise that the reward function of the AMGs is not used for training purpose but only for evaluating the score of the abstract joint policies and thus to verify if an abstract joint policy meets the optimality constraints expressed on the rewards obtained. In fact, the reward function of the concrete MG is defined as follows: 
\begin{definition} \label{def:rewardFunction}
    Each agent gets an individual reward of 1 upon collecting a flag and for reaching the goal area of the environment.
\end{definition}
% \bg{I believe that is better to give this as an equation so that we can refer to it} 
Moreover, the {\em shield} assigns a reward of 1 to each agent that completes an option and a -1 penalty to each agent that tries to perform an unsafe action, i.e. an action not allowed by the selected abstract joint policy.
%\bg{So the shield is not giving the fully coperative rewards that you mentioned before? It seems here that you are saying the shield gives the rewards to specific agents}. 
Note that we do not penalise an agent that gets caught as she is already naturally penalised by the fact that she cannot navigate the environment and get rewards anymore. Thus, even though agents have the common objective of collecting as many flags as possible before reaching the goal area of the environment, to facilitate the learning stage, they have different reward functions at the concrete level. This difference of reward function between the AMGs and the concrete MG does not impact the properties preservation as the score of the abstract joint policy is evaluated according to the atoms of the states it reaches.
%\pe{See here the reward function of the concrete MG.} \bg{I think it is not good that we refer to this function earlier but we don't explain the function until later, as soon as you want to use something you should immediately explain it} \bg{every agent receives a reward of +1 or only the agent who collects it?} \pe{Since we mentioned that we focus on fully cooperative problem, the reward function is shared with all the agents. Thus, the constraints expressed over the rewards collected are verified with this reward function. This is not the reward function used to train the agents but the reward function used to verify the constraints on the AMG.} \bg{then we should make it clearer here, are give a reference to an earlier part of the document if you already explained before. Still if this is a number to just check the constraints, I would call it score or something different, not Rewards, so that we avoid confusion}\bg{Reference where you detail this function, I saw that we defined it later, that's not a good practice}.

To evaluate AMARL, we first run the AMG construction stage to obtain the safe and optimal AMGs. In the second stage, we generate 1,000 abstract joint policies and verify our constraints on them by using the {\em Storm} model checker \cite{Storm}. Given the large number of state of our problem and the fact that the number of possible abstract joint policies increases exponentially with the number of states, it is difficult to find a policy that satisfies the expressed constraints. In the current setting, we only obtain one abstract joint policy that meets the constraints (see Table~\ref{optimal-pareto-3-agents}). Finally, we run the safe learning stage of our method and obtain the results presented in Table \ref{table:resultDRL}. Note that at test time, the shield of our method is no longer active and agents follow their learned policies in a traditional way and that during training agents always meet the safety constraints thanks to the shield. 
From results in Table \ref{table:resultDRL}, we see that the learned policies satisfy the safety constraints 
% \bg{the results only show test time, right? what about training?} \pe{Yes it is the evaluation of the final policies. I'm not sure what we can present about training? Something like the graph I had in the report of my Msc Project?} \bg{No I was thinking about just adding here some metric or reference saying that the safety constraints in training are satisfied. If you have a graph proving that from your Msc project, I suggest to add it to the supplemental material as well and leave just a reference here} 
as well as the optimality constraints defined in Table \ref{safety-constraints}. Additionally, the constraints satisfied by the learned policies closely 
%match the results returned by {\em Storm} meaning that the agents learned to perform the selected abstract joint policy but not for some rare scenarios \bg{this could lead to undesired conclusions from the reviewers, I would rather say: "... }
match the results returned by {\em Storm} meaning that the agents learned polices just marginally worse than optimal. This small divergence is potentially caused by scenarios where one or more agents get captured which were not frequent enough during the training stage for the remaining agents to learn the selected abstract joint policy. 

%\sout{In other words, it is possible that scenarios where one or more agents get captured are not frequent enough to allow the remaining agents to learn the selected abstract joint-policy. Thus, the evaluation of the final policy does not exactly match the values returned by {\em Storm} given that for some cases the agents do not complete the abstract joint-policy}. \fb{maybe this can be said more precisely. Why closely match? and why 'not for some rare scenarios'?}. \pe{Is it clearer?}

\begin{table}[t]
    \caption{Abstract joint policy's properties returned by Storm that satisfy the constraints. $P_?(\Phi)$ denotes the probability of reaching a state that satisfies $\Phi$, while $R_?(\Phi)$ denotes the expected reward obtained by reaching such a state.}
    \label{optimal-pareto-3-agents}
    \resizebox{\columnwidth}{!}{
    \begin{tabular}{@{}ccccc@{}}
    \multicolumn{5}{c}{\textbf{Optimality properties}} \\
    \toprule
    $\bm{P_{?}(\lozenge goal_{all})}$ &
    $\bm{P_{?}(\lozenge goal_1)}$ & 
    $\bm{P_{?}(\lozenge goal_2)}$ & 
    $\bm{P_{?}(\lozenge goal_3)}$ &
    $\bm{R_{?}(\lozenge end_{all})}$  \\ \midrule
    0.8393 & 1 & 0.8835 & 0.9422 & 7.8007 \\ \bottomrule
    \end{tabular}}
    \bigskip
    \resizebox{\columnwidth}{!}
    {\begin{tabular}{@{}cccc@{}}
    \multicolumn{4}{c}{\textbf{Safety properties}} \\
    \toprule
    $\bm{P_{?}(\lozenge captured_{all})}$ &
    $\bm{P_{?}(\lozenge captured_1)}$ & 
    $\bm{P_{?}(\lozenge captured_2)}$ & 
    $\bm{P_{?}(\lozenge captured_3)}$ \\ \midrule
    0 & 0 & 0.297 & 0.176 \\ \bottomrule
    \end{tabular}}
\end{table}

\begin{table}[t]
    \caption{AMARL test performance applying IDQL on the abstract joint policy from Table~\ref{optimal-pareto-3-agents}. Results are presented as the mean and standard deviation from 5 independent runs.}
    \label{table:resultDRL}
    \begin{tabular}{@{}cccc@{}}
    \toprule
    $\bm{i}$ & $\bm{P_{?}(\lozenge captured_i)}$  & $\bm{P_{?}(\lozenge goal_i)}$ & $\bm{R_{?}(\lozenge end_{i})}$ \\ \midrule
     $\bm{1}$ & 0.0 (0.0) & 0.9859 (0.271) & 1.9859 (0.0271) \\
     $\bm{2}$ & 0.1184 (0.0033) & 0.8733 (0.0195) & 3.8022 (0.022) \\
     $\bm{3}$ & 0.0523 (0.0029) & 0.8739 (0.0189) & 1.9111 (0.0297) \\ 
     $\bm{all}$ & 0.0 (0.0) & 0.8379 (0.0023) & 7.6992 (0.0553) \\
     \bottomrule
    \end{tabular}
\end{table}

In order to fairly evaluate the impact the {\em shield} has, we run a second experiment without the use of this element, whereby an episode terminates as soon as some agent reaches an unsafe state (i.e., a state that violates the joint abstract policy) and the agent that performed the unsafe action is penalised with a reward of -1. Further, to make the comparison as fair as possible, we keep the same reward function as before (See Def.~\ref{def:rewardFunction}) where additionally, agents are given a reward of 1 when completing an option and their interactions with the environment are restricted until the termination of the joint option so that we keep the {\em $T_{all}$} termination scheme. 

%\pe{Please pay attention to the above as I made modification according to the reward function.} \bg{I liked it, I just did a few very minor amends}
% Consequently, the aim of this experiment is to show how frequently agents behave unsafely w.r.t.~the selected abstract joint policy. 
The performance of the solutions learned in this experiment are provided in Table \ref{table:resultNoShield}. Contrasting these results with the ones in Table~\ref{table:resultDRL} we see that the use of the shield has no negative impact on the final performance of the agents. Nevertheless, when learning without the shield the agents reach an unsafe state in 61\% of the episodes over the 5 independent runs, while in the shield-based approach safety is always guaranteed. Finally, we also compared our method with a vanilla IDQL approach that does not take advantage of the abstraction. We observe that, in this case, the agents do not converge to an optimal solution in addition to their unsafe behaviours. We refer to Appendix \ref{appendixC} for details on this experiment.

\begin{table}[H]
    \caption{AMARL test performance after being trained without the shield on the abstract joint policy from Table \ref{optimal-pareto-3-agents}. Results are presented as in Table~\ref{table:resultDRL}.}
    \label{table:resultNoShield}
    \begin{tabular}{@{}cccc@{}}
    \toprule
    $\bm{i}$ & $\bm{P_{?}(\lozenge captured_i)}$  & $\bm{P_{?}(\lozenge goal_i)}$ & $\bm{R_{?}(\lozenge end_{i})}$ \\ \midrule
     $\bm{1}$ & 0.0 (0.0) & 0.9192 (0.0297) &  1.9044 (0.0522) \\
     $\bm{2}$ & 0.1264 (0.0258) & 0.8375 (0.0316) & 3.7555 (0.0565) \\
     $\bm{3}$ & 0.0529 (0.0043) & 0.8553 (0.0414) & 1.8902 (0.0334) \\ 
     $\bm{all}$ & 0.0 (0.0) & 0.8293 (0.0242) & 7.5501 (0.0716) \\
     \bottomrule
    \end{tabular}
\end{table}

% \begin{table}[t]
%     \caption{Results of traditional IDQL on the GFC domain with three agents. We run five experiments and present the result as follows: mean (standard deviation).}
%     \label{table:resultDRLClassic}
%     \centering
%     \begin{tabular}{@{}cccc@{}}
%     \toprule
%     & \textbf{Prob. getting caught} & \textbf{Prob. reach goal} & \textbf{Reward} \\ \midrule
%      Agent$_1$ & 0.1173 (0.0013) & 0.7059 (0.353) & 1.6459 (0.3528) \\
%      Agent$_2$ & 0.0015 (0.003) & 0.0 (0.0) &  1.398 (0.8005) \\
%      Agent$_3$ & 0.047 (0.0014) & 0.9167 (0.029) & 2.2734 (0.4171) \\ 
%      All agents & 0.0 (0.0) & 0.0 (0.0) & 5.3261 (0.7375) \\
%      \bottomrule
%     \end{tabular}
% \end{table}

%%%%%%%%%%%%%%%%%%%%%%%%%%%%%%%%%%%%%%%%%%%%%%%%%%%%%%%%%%%%%%%%%%%%%%%%

\section{Conclusion And Future Work}

In this paper we put forward 
%This work was undertaken to design 
a methodology to formally guarantee that solutions learned using deep RL algorithms in a multi-agents setting satisfy specific constraints, even during the learning stage.
%agents would act safely according to some specific safety constraints. 
By taking as starting point the ARL method proposed 
%by Mason et al.~
in \cite{AssuranceInReinforcement}, we defined the notion of abstract Markov game meant to reduce the complexity of the original problem, modelled as a Markov game. %by abstracting unnecessary features. 
The construction of the AMG 
%only requires some prior knowledge of the system and 
allows to increase the scalability of the verification method by applying model checking techniques on the abstract model. We also defined a notion of stutter bisimulation for MGs,  
%class of PCTL formulas -- 
 which we proved to be adapt to preserve the weak fragment of PCTL. We thus established the conditions an AMG has to satisfy to guarantee that true properties, expressed in wPCTL, are preserved in the corresponding MG.

Furthermore, in order to keep the learning stage efficient, we defined a relaxation of stutter bisimulation to allow agents to act as independently as possible. Accordingly, we introduced the classes of wPCTL formulae that are 
%guaranteed to be 
preserved 
%in the original MG 
when the learning stage takes advantage of this relaxation. Therefore, 
%in addition to the extension of the ARL method to multi-agent settings, 
we provided formal guarantees of property preservation between an AMG and its corresponding MG. This contribution, unlike the ARL method, allows us to ensure the safe behaviours of agents both at training and testing time.
%during the learning stage and when following the learned policy. 
Moreover, we provided an algorithm to generate the bisimilar AMG automatically.
%so that it satisfies the properties expressed by the stutter bisimulation. 
Finally, our AMARL method allows for choosing beforehand the policy to be learnt among a set of safe solutions, thus permitting to select the most adapted 
%\fb{in what sense adapted?} \pe{In the sense that for a specific problem, one would prefer having a solution as safe as possible while others would prefer to have a solution that meets the constraints but that try to optimize a specific criterion} 
%\fb{Ok, but how is the trade-off between safety and optimality managed? is safety prioritized over optimality? my point is that we have to explain what we mean by 'adapted'.} \pe{I see, then I would say that it depends on the objective of the engineerer. For some problems, safety may be prioritized while for other problem we will want to prioritze the optimality with minimal safety insurance. I have adapted the text accordingly, is it better?}
solution for the problem regarding both safety and optimality, depending on the application at hand.

The empirical evaluation we ran on the GFC domain showed that the safety constraints were always satisfied, and that the solution learned using DRL almost always converged to the optimal solution with respect to the selected abstract joint policy, and that it improves on the solution found using the same DRL algorithm without AMARL. The experiments also showed that in addition to ensuring agents safety, the shield of our method had no negative impact and even improved by a bit the agents' final performance in comparison to AMARL without the shield.
%that the solutions found using the AMARL method surpass those found by using ARL.
%method with a single agent. 

Future work on the AMARL method would include increasing its scalability,
%of the method 
also through a more efficient implementation of the procedure to automatically generate the abstract MG. 
%Moreover, way to improve the discovery of policies that satisfies the constraints need to be proposed. An idea that was proposed in \cite{AssuranceInReinforcement} is to use genetic algorithms \cite{gerasimou2015search}. 
A natural continuation of this work would deal with different termination scheme, as well as support competitive and mixed games. Finally, a further work is needed to 
%assess the 
apply AMARL to a wider range of problems as well as 
%or could try to express 
a broader class of PCTL contraints.

%Other directions for future work would include taking a different approach than the one we took concerning the verification of the constraints in the abstract model. Instead of generating abstract policies, an approach similar to the one proposed by Pathak et al. \cite{VerificationRepairControlPolicies} would verify the complete abstract system and remove transitions that violate the constraints. Such transitions could be easily identified as the model checker would return a counter-example. This kind of approach would allow using different types of abstraction, and thus different classes of PCTL formulas could be preserved between the abstract and concrete model.

\bibliographystyle{ACM-Reference-Format} 
\bibliography{sample}

\newpage 

\appendix

\section{Proofs of Theorems and Lemmas} \label{appendixA}

In order to prove our Lemmas and Theorems, we are first required to introduce the following notations.

\paragraph{Notation}  For infinite path fragments $w_i = s_{0,i}s_{1,i}s_{2,i}..., i=1,2$, we say that $w_1$ and $w_2$ are bisimilar, denoted $w_1 \simeq w_2$, iff both paths can be divided into segments $s_{j_r,1}s_{j_r+1,1}s_{j_r+2,1}...s_{j_{r+1},1}$ and $s_{k_r,2}s_{k_r+1,2}s_{k_r+2,2}$ $...s_{k_{r+1},2}$ respectively with $r=1,2,3...$, $0 = j_0 < j_1 < j_2 < ...$, $0 = k_0 < k_1 < k_2 < ...$ and that are statewise bisimilar. 

Similarly, we sat that $w_1$ and $w_2$ are partially bisimilar, denoted $w_1 \simeq_r w_2$, iff both paths can be divided into segments $s_{j_r,1}s_{j_r+1,1}$ $s_{j_r+2,1}$ $...s_{j_{r+1},1}$ and $s_{k_r,2}s_{k_r+1,2}s_{k_r+2,2}...s_{k_{r+1},2}$ respectively with $r=1,2,3...$, $0 = j_0 < j_1 < j_2 < ...$, $0 = k_0 < k_1 < k_2 < ...$, where for every $r$, $s_{j_r, 1} \simeq_r s_{k_r, 2}$ and every state between $s_{j_r, 1}$ and $s_{j_{r+1}, 1}$ and between $s_{k_r, 1}$ and $s_{j_{k+1}, 1}$, respect the relaxed branching condition.

\subsection{Proof of Lemma \ref{lemma:divergence-sensitive} and Theorem \ref{weakBisimulationTheoremMG}}

The proof of Theorem \ref{weakBisimulationTheoremMG} is done by induction over the structure of $\Phi$ similarly to what is done in \cite{principlesmodelchecking} for the proof of preservation properties of stutter bisimulation for deterministic systems. However, we first need to introduce the following definition and lemma.

\begin{definition} \label{def:pathsbisimilarMG}
    Let $M_1$ and $M_2$ be two Markov games. For each $w \in Paths(M)$, $[w]_{\simeq}$ denotes the equivalence class of path $w$, i.e., $[w]_{\simeq} = \{w' \in Paths(M) \mid w \simeq w'\}$. $Paths(M)$ denotes the set of all paths available in $M$, i.e. $Paths(M) = \bigcup_{s \in S} Paths(s)$ where $Paths(s)$ denotes the set of all paths available from $s$. We write $P_\pi([w]_{\simeq}) = \sum_{w \in [w]_{\simeq}} P_\pi(w)$ to denotes the probability that a path $w \in [w]_{\simeq}$ occurs when following the policy $\pi$. We write $W/M$ to denotes the set consisting of all equivalence classes of path of $M$.
\end{definition}

\begin{lemma} \label{lemma:probabilitypathsMG}
    Let $M_1$ and $M_2$ be two Markov games over $AP$ with labelling functions $L_i: S_i \rightarrow 2^{AP}$ for $M_1$ and $M_2$. Given $s_1 \in S_1$ and $s_2 \in S_2$, if $s_1 \simeq s_2$, then: 
    \begin{itemize}
        \item For all $\pi_1, w_1$ such that $P_{\pi_1}([w_1]_{\simeq}) > 0$, there exists $\pi_2, w_2$ such that $w_1 \simeq w_2$ and  $P_{\pi_1}([w_1]_{\simeq}) = P_{\pi_2}([w_2]_{\simeq})$.
        \item For all $\pi_2, w_2$ such that $P_{\pi_2}([w_2]_{\simeq}) > 0$, there exists $\pi_1,w_1$ such that $w_1 \simeq w_2$  and $P_{\pi_1}([w_1]_{\simeq}) = P_{\pi_2}([w_2]_{\simeq})$.
    \end{itemize}
    where $w_1 \in Paths(s_1)$ and $w_2 \in Paths(w_2)$.
    
    In words, the first item states that for every policy $\pi_1$ and every path $w_1$ that belongs to the set of all possible paths that start in $s_1$ and that can occur when following the policy $\pi_1$, there exist a policy $\pi_2$ and a path $w_2$ such that $w_1 \simeq w_2$ and the probability that a bisimilar path of $w_1$ in $M_1$ occurs is equal to the probability that a bisimilar path of $w_2$ in $M_2$ occurs. The second item is its symmetric counterpart.
\end{lemma}

Notice that Lemma \ref{lemma:divergence-sensitive} is a special case of Lemma \ref{lemma:probabilitypathsMG} where the possible policies are restricted to policies that always remain in the same equivalence class. Therefore, by proving Lemma \ref{lemma:probabilitypathsMG}, we also prove Lemma \ref{lemma:divergence-sensitive}.

\begin{proof}
    The proof of the first statement of Lemma \ref{lemma:probabilitypathsMG} directly follows from our definition of bisimulation and is done by induction. The proof of the second statement is omitted as by symmetry, is similar.
     \paragraph{Induction basis.} The induction basis (i=0) is straightforward as it only consists of the first state of the path. By definition we have that $s_{0,1} \simeq s_{0,2}$ and therefore for the path fragment $w_1=s_{0,1}$ we have a path fragment $w_2=s_{0,2}$ with both probability 1.
     
     \paragraph{Induction step.} Assume $i \geq 0$ and that the path fragments:
     \begin{equation*}
        \begin{split}
            w_1 & = s_{0,1}s_{1,1}s_{2,1}...s_{i,1}(1) \\
            w_2 & = s_{0,2}u_{0,0}...u_{0,n_0}s_{1,2}u_{1,0}...u_{1,n_1}s_{2,2}...s_{i,2}(2)
        \end{split}
    \end{equation*}
    are already constructed and that $P_{\pi_1}([w_1]_{\simeq}) = P_{\pi_2}([w_2]_{\simeq})$
    
    According to the definition of bisimulation, since $s_{i,1} \simeq s_{i,2}$, for every possible transition from $s_{i,1}$ to some states $s_{i+1,1}$ following $\pi_1$, there exists $k \geq 1$ finite path fragments of the form $s_{i,2},u_{i,0}...u_{i,n_i}s_{i+1,2}$ following $\pi_2$ such that:
    \begin{equation*}
        s_{i+1,1} \simeq s_{i+1, 2}, s_{i, 2} \simeq u_{i,0} \simeq ... \simeq u_{i,n_i}
    \end{equation*}
    and for each possible $s_{i+1, 1}$, the probability of reaching each equivalence class $[s_{i+1, 1}]_\mathcal{E}$ from $s_{i,1}$, denoted $P_{\pi_1}(s_{i,1}, [s_{i+1, 1}]_\mathcal{E})$, is equal to the probability of reaching each equivalence class $[s_{i+1, 2}]_\mathcal{E}$ from $s_{i,2}$, denoted $P_{\pi_2}(s_{i21}, [s_{i+1, 2}]_\mathcal{E})$. By appending each possible state $s_{i+1,1}$ to the path (1) and each path fragments $s_{i,2},u_{i,0}...u_{i,n_i}s_{i+1,2}$ to the path (2), we obtain a set of paths that meet the desired conditions, i.e., each paths resulting of the concatenation of (1) and states $s_{i+1_1}$ that belongs to the same equivalence class, will be bisimilar by definition and each paths resulting of the concatenation of (2) and each path fragments $s_{i,2},u_{i,0}...u_{i,n_i}s_{i+1,2}$ with $s_{i+1,2}$ that belongs to the same equivalence class, will also be bisimilar. Thus, for each $w'_1$ resulting from the concatenation with (1), we have that $P_{\pi_1}([w'_1]_{\simeq}) = P_{\pi_1}([w_1]_{\simeq}) \cdot P_{\pi_1}(s_{i,1}, [s_{i+1, 1}]_\mathcal{E})$ and for each $w'_2$ resulting from the concatenation with (2), we have that $P_{\pi_2}([w'_2]_{\simeq}) = P_{\pi_2}([w_2]_{\simeq}) \cdot P_{\pi_2}(s_{i,2}, [s_{i+1, 2}]_\mathcal{E})$, thus for each $w'_1$ there exist a $w'_2$ such that $w'_1 \simeq w'_2$ and therefore $P_{\pi_1}([w'_1]_{\simeq}) = P_{\pi_2}([w'_2]_{\simeq})$.
\end{proof}

We now provide proof of Theorem \ref{weakBisimulationTheoremMG}.

\begin{proof}
     Let $M_1 \simeq M_2$ and let $\mathcal{E} \subseteq S_1 \cup S_2$ be the equivalence relation between them. Let $s_1 \in S_1, s_2 \in S_2$ and $w_1 \in Paths(M_1), w_2 \in Paths(M_2)$. We have:
     
     \begin{enumerate}
         \item If $s_1 \simeq s_2$ according to $\mathcal{E}$, then for any wPCTL formula $\Phi$: $s_1 \models \Phi \iff s_2 \models \Phi$.
         \item If $w_1 \simeq w_2$, then for any wPCTL path formula $\varphi$: $w_1 \models \varphi \iff w_2 \models \varphi$.
     \end{enumerate}
     
     \paragraph{Induction basis.} Let $s_1 \simeq s_2$. For $\Phi = true$, proposition (1) clearly holds. Given that $L_1(s_1) = L_2(s_2)$, we have that for $\Phi = a \in AP$:
     \begin{center}
         $s_1 \models a \iff a \in L_1(s_1) \iff a \in L_2(s_2) \iff s_2 \models a$.
     \end{center}
     
     \paragraph{Induction step.}
     Assume $\Phi_1, \Phi_2, \Phi_3, \Phi_4$ are wPCTL state formulae for which proposition (1) holds. Let $s_1 \simeq s_2$.
     
     \textbf{Case 1: $\Phi = \Phi_1 \land \Phi_2$.} The application of the induction hypothesis for $\Phi_1$ and $\Phi_2$ gives:
     \begin{equation*}
        \begin{split}
            s_1 \models \Phi_1 \land \Phi_2 & \iff s_1 \models \Phi_1 \text{ and } s_1 \models \Phi_2 \\ 
            & \iff s_2 \models \Phi_1 \text{ and } s_2 \models \Phi_2 \\ 
            & \iff s_2 \models \Phi_1 \land \Phi_2.\\ 
         \end{split}
    \end{equation*}
    
    \textbf{Case 2: $\Phi = \neg \Phi_3$.} The application of the induction hypothesis for $\Phi_3$ gives:
    \begin{equation*}
        \begin{split}
            s_1 \models \neg \Phi_3 & \iff s_1 \not\models \Phi_3 \\ 
            & \iff s_2 \not\models \Phi_3 \\ 
            & \iff s_2 \models \neg \Phi_3. \\ 
        \end{split}
    \end{equation*}
    
    \textbf{Case 3: $\Phi = P_{\sim p}(\varphi)$.} 
    We only prove the case where $\Phi = P_{\geq p}(\varphi)$ as the other cases are similar to this one.
    We first prove by contraposition that $s_1 \models \Phi \implies s_2 \models \Phi$: Assume that $s_2 \not\models \Phi$. According to the satisfaction relation of wPCTL, we have that if $s_2 \not\models \Phi$, it means that there exists a policy $\pi_2$ of $M_2$ such that:
    \begin{equation*}
        \sum_{[w_2]_{\simeq} \in W/M_2, w_2 \in Paths(s_2), w_2 \models \varphi} P_{\pi_2}([w_2]_{\simeq}) < p
    \end{equation*}
    and therefore for this policy $\pi_2$, there exists a policy $\pi_1$ such that for every $w_2 \in Paths(s_2)$ there exist a $w_1 \in Paths(s_1)$ such that $P_{\pi_2}([w_2]_{\simeq}) = P_{\pi_1}([w_1]_{\simeq})$ and $w_1 \simeq w_2$ by Lemma \ref{lemma:probabilitypathsMG}. From the induction hypothesis (2) it follows that if $w_1 \simeq w_2$, $w_2 \models \varphi \iff w_1 \models \varphi$ and therefore according to lemma \ref{lemma:probabilitypathsMG}, if $s_2 \not\models \Phi$, then $s_1 \not\models \Phi$. Thus we proved that $s_2 \not \models \Phi \implies s_1 \not \models \Phi$, that is, $s_1 \models \Phi \implies s_2 \models \Phi$ by contraposition. It remains to prove that $s_2 \models \Phi \implies s_1 \models \Phi$ but we omit the details for symmetry reason.
    
    \textbf{Case 4: $\Phi = R_{\sim r}(\lozenge \Phi_4)$.} The proof for this case is similar to the proof of Case 3, and we, therefore, omit the details for simplicity. The only non-trivial part of the proof for this case is that is is necessary to notice that $\lozenge \Phi_4$ is equal to the path formulae $\varphi = true$ $U$ $\Phi_4$ and thus from the induction hypothesis (2) it follows that if $w_1 \simeq w_2$, $w_2 \models \varphi \iff w_1 \models \varphi$.
    
    We now prove statement (2). Assume claim (1) holds for wPCTL state formulae $\Phi_1$ and $\Phi_2$. Let $w_1 \simeq w_2$. We write $w_i[...j]$ to denotes the prefix $s_{0,i}s_{1,i}s_{2,i}...s_{j-1, i}$ of $w_i$. The only case we need to prove for wPCTL path formula is the case where $\varphi = \Phi_1 U \Phi_2$.
    
    The induction hypothesis for $\Phi_1$ and $\Phi_2$ and the path fragments $w_i, i=1,2$ gives:
    \begin{tabbing}
        $w_1 \models \Phi_1 U \Phi_2$ \= $\iff$ \= there exists an index $j_r \in \mathbb{N}$ with \\
        \> \> $w_1[j_r] \models \Phi_2$  and \\
        \> \> $w_1[...j_r] \models \Phi_1$ \\
        \> $\iff$ \> there exists ans index $k_r \in \mathbb{N}$ with \\
        \> \> $w_2[k_r] \models \Phi_2$  and \\
        \> \> $w_2[...k_r] \models \Phi_1$ \\
        \> $\iff$ \> $w_2 \models \Phi_1 U \Phi_2$
    \end{tabbing}
    where $r \in \mathbb{N}^*$. The above clearly hold by definition of the stutter bisimulation.
\end{proof}

\subsection{Proof of Theorem \ref{AgentAPrelaxedStutterBisimulationTheoremMG}}

The proof of Theorem \ref{AgentAPrelaxedStutterBisimulationTheoremMG} of our paper is similar to the proof of Theorem \ref{weakBisimulationTheoremMG} but for some details. We thus adapt Definition.~\ref{def:pathsbisimilarMG} and Lemma~\ref{lemma:probabilitypathsMG} for the relaxed version of bisimulation.

\begin{definition} \label{def:relaxedpathsbisimilarMG}
     Let $M_1$ and $M_2$ be two Markov games. For each $w \in Paths(M)$, $[w]_{\simeq_r}$ denotes the partial equivalence class of path $w$, i.e., $[w]_{\simeq_r} = \{w' \in Paths(M) \mid w \simeq_r w'\}$. We write $P_\pi([w]_{\simeq_r}) = \sum_{w \in [w]_{\simeq_r}} P_\pi(w)$ to denotes the probability that a path $w \in [w]_{\simeq_r}$ occurs when following the policy $\pi$. We write $W_r/M$ to denotes the set consisting of all partial equivalence classes of path of $M$.
\end{definition}

\begin{lemma} \label{lemma:relaxedprobabilitypathsMG}
     Let $M_1$ and $M_2$ be two Markov games over $AP$ with labelling functions $L_i: S_i \rightarrow 2^{AP}$ for $M_1$ and $M_2$. Given $s_1 \in S_1$ and $s_2 \in S_2$, if $s_1 \simeq_r s_2$, then: 
    \begin{itemize}
        \item For all $\pi_1, w_1$ such that $P_{\pi_1}([w_1]_{\simeq_r}) > 0$, there exists $\pi_2,w_2$ such that $w_1 \simeq_r w_2$ and  $P_{\pi_1}([w_1]_{\simeq_r}) = P_{\pi_2}([w_2]_{\simeq_r})$.
        \item For all $\pi_2, w_2$ such that $P_{\pi_2}([w_2]_{\simeq_r}) > 0$, there exists $\pi_1,w_1$ such that $w_1 \simeq_r w_2$  and $P_{\pi_1}([w_1]_{\simeq_r}) = P_{\pi_2}([w_2]_{\simeq_r})$.
    \end{itemize}
    where $w_1 \in Paths(s_1)$ and $w_2 \in Paths(w_2)$.
    
    In words, the first item states that for every policy $\pi_1$ and every path $w_1$ that belongs to the set of all possible paths that start in $s_1$ and that can occur when following the policy $\pi_1$, there exist a policy $\pi_2$ and a path $w_2$ such that $w_1 \simeq_r w_2$ and the probability that a partially bisimilar path of $w_1$ in $M_1$ occurs is equal to the probability that a partially bisimilar path of $w_2$ in $M_2$ occurs. The second item is its symmetric counterpart.
\end{lemma}

The proof of Lemma \ref{lemma:relaxedprobabilitypathsMG} is a straightforward adaptation of the proof of Lemma \ref{lemma:probabilitypathsMG} where only the notations change. The reason is that the relaxation of the branching condition does not impact the probability of reaching the next equivalence classes. We, therefore, omit the proof here and only provide the proof for Theorem \ref{AgentAPrelaxedStutterBisimulationTheoremMG}.

\begin{proof}
    We first prove statement (1) of Theorem \ref{AgentAPrelaxedStutterBisimulationTheoremMG} by induction over the structure of $\Phi$ in a similar way than the proof of Theorem \ref{weakBisimulationTheoremMG}.
    
    Let $M_1 \simeq_r M_2$ and let $\mathcal{E} \subseteq S_1 \cup S_2$ be the equivalence relation between them. Let $s_1 \in S_1, s_2 \in S_2$ and $w_1 \in Paths(M_1), w_2 \in Paths(M_2)$. We have:
    \begin{enumerate}[(a)]
        \item If $s_1 \simeq_r s_2$ according to $\mathcal{E}$, then for any wPCTL formula $\Phi$ over $AP_{agent_i}$: $s_1 \models \Phi \iff s_2 \models \Phi$.
        \item If $w_1 \simeq_r w_2$, then for any wPCTL path formula $\varphi$ over $AP_{agent_i}$: $w_1 \models \varphi \iff w_2 \models \varphi$.
    \end{enumerate}
     
    \paragraph{Induction basis.} Let $s_1 \simeq_r s_2$. For $\Phi = true$, proposition (a) clearly holds. Given that $L_1(s_1) = L_2(s_2)$, we have that for $\Phi = a \in AP$:
    \begin{center}
        $s_1 \models a \iff a \in L_1(s_1) \iff a \in L_2(s_2) \iff s_2 \models a$.
    \end{center}
     
    \paragraph{Induction step.}
     Assume $\Phi_1, \Phi_2, \Phi_3, \Phi_4$ are wPCTL state formulae over $AP_{agent_i}$ for which proposition (a) holds. Let $s_1 \simeq_r s_2$.
     
    \textbf{Case 1: $\Phi = \Phi_1 \land \Phi_2$.} The application of the induction hypothesis for $\Phi_1$ and $\Phi_2$ gives:
    \begin{equation*}
        \begin{split}
            s_1 \models \Phi_1 \land \Phi_2 & \iff s_1 \models \Phi_1 \text{ and } s_1 \models \Phi_2 \\ 
            & \iff s_2 \models \Phi_1 \text{ and } s_2 \models \Phi_2 \\ 
            & \iff s_2 \models \Phi_1 \land \Phi_2.\\ 
         \end{split}
    \end{equation*}

    \textbf{Case 2: $\Phi = \neg \Phi_3$.} The application of the induction hypothesis for $\Phi_3$ gives:
    \begin{equation*}
        \begin{split}
            s_1 \models \neg \Phi_3 & \iff s_1 \not\models \Phi_3 \\ 
            & \iff s_2 \not\models \Phi_3 \\ 
            & \iff s_2 \models \neg \Phi_3. \\ 
        \end{split}
    \end{equation*}
    
    \textbf{Case 3: $\Phi = P_{\sim p}(\varphi)$.} 
    We only prove the case where $\Phi = P_{\geq p}(\varphi)$ as the other cases are similar to this one.
    We first prove by contraposition that $s_1 \models \Phi \implies s_2 \models \Phi$: Assume that $s_2 \not\models \Phi$. According to the satisfaction relation of wPCTL, we have that if $s_2 \not\models \Phi$, it means that there exist a policy $\pi_2$ of $M_2$ such that:
    \begin{equation*}
        \sum_{[w_2]_{\simeq_r} \in W_r/M_2, w_2 \in Paths(s_2), w_2 \models \varphi} P_{\pi_2}([w_2]_{\simeq_r}) < p
    \end{equation*}
    and therefore for this policy $\pi_2$, there exists a policy $\pi_1$ such that for every $w_2 \in Paths(s_2)$ there exist a $w_1 \in Paths(s_1)$ such that $P_{\pi_2}([w_2]_{\simeq_r}) = P_{\pi_1}([w_1]_{\simeq_r})$ and $w_1 \simeq_r w_2$ by Lemma \ref{lemma:relaxedprobabilitypathsMG}. From the induction hypothesis (b) it follows that if $w_1 \simeq_r w_2$, $w_2 \models \varphi \iff w_1 \models \varphi$ and therefore according to lemma \ref{lemma:relaxedprobabilitypathsMG}, if $s_2 \not\models \Phi$, then $s_1 \not\models \Phi$. Thus we proved that $s_2 \not \models \Phi \implies s_1 \not \models \Phi$, that is, $s_1 \models \Phi \implies s_2 \models \Phi$ by contraposition. It remains to prove that $s_2 \models \Phi \implies s_1 \models \Phi$ but we omit the details for symmetry reason.
    
    \textbf{Case 4: $\Phi = R_{\sim r}(\lozenge \Phi_4)$.} The proof for this case is similar to the proof of Case 3, and we, therefore, omit the details for simplicity. The only non-trivial part of the proof for this case is that is is necessary to notice that $\lozenge \Phi_4$ is equal to the path formulae $\varphi = true$ $U$ $\Phi_4$ and thus from the induction hypothesis (b) it follows that if $w_1 \simeq_r w_2$, $w_2 \models \varphi \iff w_1 \models \varphi$.
    
    We now prove statetement (b). Assume claim (a) holds for wPCTL state formulae $\Phi_1$ and $\Phi_2$ over $AP_{agent_i}$. Let $w_1 \simeq_r w_2$. We write $w_i[...j]$ to denotes the prefix $s_{0,i}s_{1,i}s_{2,i}...s_{j-1, i}$ of $w_i$. The only case we need to prove for wPCTL path formula over $AP_{agent_i}$ is the case where $\varphi = \Phi_1 U \Phi_2$.
    
    The induction hypothesis for $\Phi_1$ and $\Phi_2$ and the path fragments $w_i, i=1,2$ gives:
    \begin{tabbing}
        $w_1 \models \Phi_1 U \Phi_2$ \= $\iff$ \= there exists an index $j_r \in \mathbb{N}$ with \\
        \> \> $w_1[j_r] \models \Phi_2$  and \\
        \> \> $w_1[...j_r] \models \Phi_1$ \\
        \> $\iff$ \> there exists ans index $k_r \in \mathbb{N}$ with \\
        \> \> $w_2[k_r] \models \Phi_2$  and \\
        \> \> $w_2[...k_r] \models \Phi_1$ \\
        \> $\iff$ \> $w_2 \models \Phi_1 U \Phi_2$
    \end{tabbing}
    where $r \in \mathbb{N}^*$.\\\\
    The above notation holds because of the definition of the relaxed stutter bisimulation requires that if one AP $a \in AP_{agent_i}$ changes during a path $w$, all the atomic propositions of $agent_i$ have to change simultaneously. It is therefore easy to see that if there exists such an index for $w_1$, there also exists an index for $w_2$ that fulfils the required conditions.
    \balance
    We now prove statement (2) of Theorem 3.7 in a similar way than above.
    
    Let $M_1 \simeq M_2$ and let $\mathcal{E} \subseteq S_1 \cup S_2$ be the equivalence relation between them. Let $s_1 \in S_1, s_2 \in S_2$ and $w_1 \in Paths(M_1), w_2 \in Paths(M_2)$. We have:
     \begin{enumerate}[(a)]
         \item If $s_1 \simeq s_2$ according to $\mathcal{E}$, then for any wPCTL formula $\Phi$: $s_1 \models \Phi \iff s_2 \models \Phi$.
         \item If $w_1 \simeq w_2$, then for any wPCTL path formula $\varphi = (true$ $U$ $\Phi)$: $w_1 \models \varphi \iff w_2 \models \varphi$.
         \item If $w_1 \simeq w_2$, and $\Phi_1$, $\Phi_2$ only contain the same AP $a$, then for any wPCTL path formula $\varphi = (\Phi_1$ $U$ $\Phi_2)$: $w_1 \models \varphi \iff w_2 \models \varphi$.
     \end{enumerate}
     
     The proof of statement (2) is the same as the proof of statement (1) at the exception of the proof on path formulae. This is due to the fact that the relaxation of the branching condition only impacts the paths of options of the stutter bisimulation. For this reason, we omit details of the statement (a) and only provide the proof of statements (b) and (c).
     
    Assume claim (a) holds for wPCTL state formula $\Phi$. Let $w_1 \simeq_r w_2$. We write $w_i[...j]$ to denotes the prefix $s_{0,i}s_{1,i}s_{2,i}...s_{j-1, i}$ of $w_i$. We first prove the case where $\varphi = true$ $U$ $\Phi$.

    The induction hypothesis for $\Phi$ and the path fragments $w_i, i=1,2$ gives:
    \begin{tabbing}
        $w_1 \models true$ $U$ $\Phi$ \= $\iff$ \= there exists an index $j_r \in \mathbb{N}$ with \\
        \> \> $w_1[j_r] \models \Phi$  and \\
        \> \> $w_1[...j_r] \models$ $true$ \\
        \> $\iff$ \> there exists ans index $k_r \in \mathbb{N}$ with \\
        \> \> $w_2[k_r] \models \Phi$  and \\
        \> \> $w_2[...k_r] \models$ $true$ \\
        \> $\iff$ \> $w_2 \models true$ $U$ $\Phi$
    \end{tabbing}
    where $r \in \mathbb{N}^*$.
    
    It is easy to see that the properties above hold as it directly follows from the definition of partially stutter bisimilar paths.

    It then remains to prove statement (c). Assume claim (a) holds for wPCTL formulae $\Phi_1$, $\Phi_2$ that only contains the same atomic proposition $a$. The induction hypothesis for $\Phi_1$ and $\Phi_2$ and the path fragments $w_i, i=1,2$ gives:
    \begin{tabbing}
        $w_1 \models \Phi_1 U \Phi_2$ \= $\iff$ \= there exists an index $j_r \in \mathbb{N}$ with \\
        \> \> $w_1[j_r] \models \Phi_2$  and \\
        \> \> $w_1[...j_r] \models \Phi_1$ \\
        \> $\iff$ \> there exists ans index $k_r \in \mathbb{N}$ with \\
        \> \> $w_2[k_r] \models \Phi_2$  and \\
        \> \> $w_2[...k_r] \models \Phi_1$ \\
        \> $\iff$ \> $w_2 \models \Phi_1 U \Phi_2$
    \end{tabbing}
    where $r \in \mathbb{N}^*$.

    Again, it is easy to see that these properties hold as the relaxed branching condition requires that if an AP changes along a path, this AP has to take the value it has in the last state of the path and has to keep it until it reaches the last state of the path. Otherwise, the AP keeps the value of the initial state.
    
    Finally, note that the case where the path formula is equal to a state is expressed by statement (a) and no additional proof is needed for this case.
\end{proof}

\section{Training details} \label{appendixB}

For all our experiments we use the following hyperparameters:
\begin{itemize}
    \item Batch size: 64 
    \item Replay buffer size: 1,000,000
    \item Episode length: 1,000
    \item Epsilon (decay): 1 (0.99988), decay to 0.1
    \item Learning Rate: 0.0001
    \item Discount factor: 0.95
    \item Number of episodes: 20,000
    \item Target network updated every 1,000 Q-network updates. We train the network 5 times every 10 steps.
\end{itemize}

\section{Vanilla IDQL experiment} \label{appendixC}

We provide in Table \ref{table:resultDRLClassic} the results of vanilla IDQL on our GFC domain. This experiment uses the same reward function, architecture and hyperparameters than the experiments presented in our main paper. 

\begin{table}[t] 
    \caption{Results of vanilla IDQL on the GFC domain with three agents. Results are presented as the mean and standard deviation from 5 independent runs.}
    \label{table:resultDRLClassic}
    \centering
    \begin{tabular}{@{}cccc@{}}
        \toprule
        $\bm{i}$ & $\bm{P_{?}(\lozenge captured_i)}$  & $\bm{P_{?}(\lozenge goal_i)}$ & $\bm{R_{?}(\lozenge end_{i})}$ \\ \midrule
        $\bm{1}$ & 0.1173 (0.0013) & 0.7059 (0.353) & 1.6459 (0.3528) \\
        $\bm{2}$ & 0.0015 (0.003) & 0.0 (0.0) &  1.398 (0.8005) \\
        $\bm{3}$ & 0.047 (0.0014) & 0.9167 (0.029) & 2.2734 (0.4171) \\ 
        $\bm{all}$ & 0.0 (0.0) & 0.0 (0.0) & 5.3261 (0.7375) \\
        \bottomrule
    \end{tabular}
\end{table}

\begin{table}[t]
    \caption{Performance of the random policy when applying our safe exploration method vs. no restriction.}
    \label{table:randomSafePolicy}
    \resizebox{\columnwidth}{!}{
    \begin{tabular}{@{}ccc@{}}
    \toprule
     & $\bm{P_{?}(\lozenge captured_i)}$ \textbf{with shield} & $\bm{P_{?}(\lozenge captured_i)}$ \textbf{ without shield} \\ \midrule
     Agent$_1$ & 0.0 & 0.4851  \\
     Agent$_2$ & 0.1039 & 0.4601 \\
     Agent$_3$ & 0.0743 & 0.5185 \\ 
     All agents & 0.0 & 0.1154 \\
     \bottomrule
    \end{tabular}}
\end{table}

This experiment shows that vanilla IDQL is not able to converge toward an optimal policy and does not satisfy the optimality constraints. Moreover, even though the results of Table \ref{table:resultDRLClassic} satisfy the safety constraints, this is mainly due to the fact that the leanred policies of the agents do not try to collect all the flags. 

Finally, to show that the shield of our method always ensure the satisfaction of the safety constraints even during training time and thus during random exploration of the environment we evaluate the performance of the agents when following a random policy. We also repeat this experiment without the use of the shield and shows that agents violate the safety constraints. We summarise the results of these two experiments in Table \ref{table:randomSafePolicy}.

%%%%%%%%%%%%%%%%%%%%%%%%%%%%%%%%%%%%%%%%%%%%%%%%%%%%%%%%%%%%%%%%%%%%%%%%

%%% The following command should be issued somewhere in the first column 
%%% of the final page of your paper.
%\balance
%%% The acknowledgments section is defined using the "acks" environment
%%% (rather than an unnumbered section). The use of this environment 
%%% ensures the proper identification of the section in the article 
%%% metadata as well as the consistent spelling of the heading.

% \begin{acks}
% If you wish to include any acknowledgments in your paper (e.g., to 
% people or funding agencies), please do so using the `\texttt{acks}' 
% environment. Note that the text of your acknowledgments will be omitted
% if you compile your document with the `\texttt{anonymous}' option.
% \end{acks}

%%%%%%%%%%%%%%%%%%%%%%%%%%%%%%%%%%%%%%%%%%%%%%%%%%%%%%%%%%%%%%%%%%%%%%%%

%%% The next two lines define, first, the bibliography style to be 
%%% applied, and, second, the bibliography file to be used.

%\bibliographystyle{ACM-Reference-Format} 
%\bibliography{sample}

%%%%%%%%%%%%%%%%%%%%%%%%%%%%%%%%%%%%%%%%%%%%%%%%%%%%%%%%%%%%%%%%%%%%%%%%

\end{document}